\PassOptionsToPackage{unicode}{hyperref}
\PassOptionsToPackage{hyphens}{url}
\PassOptionsToPackage{dvipsnames,svgnames,x11names}{xcolor}
\documentclass[
]{article}
\usepackage{xcolor}
\usepackage{amsmath,amssymb}
\usepackage[T1]{fontenc}
\usepackage[utf8]{inputenc}
\usepackage{textcomp} % provide euro and other symbols

\usepackage{lmodern}
\makeatletter
\@ifundefined{KOMAClassName}{% if non-KOMA class
  \IfFileExists{parskip.sty}{%
    \usepackage{parskip}
  }{% else
    \setlength{\parindent}{0pt}
    \setlength{\parskip}{6pt plus 2pt minus 1pt}}
}{% if KOMA class
  \KOMAoptions{parskip=half}}
\makeatother
\makeatletter
\ifx\paragraph\undefined\else
  \let\oldparagraph\paragraph
  \renewcommand{\paragraph}{
    \@ifstar
      \xxxParagraphStar
      \xxxParagraphNoStar
  }
  \newcommand{\xxxParagraphStar}[1]{\oldparagraph*{#1}\mbox{}}
  \newcommand{\xxxParagraphNoStar}[1]{\oldparagraph{#1}\mbox{}}
\fi
\ifx\subparagraph\undefined\else
  \let\oldsubparagraph\subparagraph
  \renewcommand{\subparagraph}{
    \@ifstar
      \xxxSubParagraphStar
      \xxxSubParagraphNoStar
  }
  \newcommand{\xxxSubParagraphStar}[1]{\oldsubparagraph*{#1}\mbox{}}
  \newcommand{\xxxSubParagraphNoStar}[1]{\oldsubparagraph{#1}\mbox{}}
\fi
\makeatother

\usepackage{longtable,booktabs,array}
\usepackage{calc} % for calculating minipage widths
\usepackage{etoolbox}
\makeatletter
\patchcmd\longtable{\par}{\if@noskipsec\mbox{}\fi\par}{}{}
\makeatother
\IfFileExists{footnotehyper.sty}{\usepackage{footnotehyper}}{\usepackage{footnote}}
\makesavenoteenv{longtable}
\usepackage{graphicx}
\makeatletter
\newsavebox\pandoc@box
\newcommand*\pandocbounded[1]{% scales image to fit in text height/width
  \sbox\pandoc@box{#1}%
  \Gscale@div\@tempa{\textheight}{\dimexpr\ht\pandoc@box+\dp\pandoc@box\relax}%
  \Gscale@div\@tempb{\linewidth}{\wd\pandoc@box}%
  \ifdim\@tempb\p@<\@tempa\p@\let\@tempa\@tempb\fi% select the smaller of both
  \ifdim\@tempa\p@<\p@\scalebox{\@tempa}{\usebox\pandoc@box}%
  \else\usebox{\pandoc@box}%
  \fi%
}
\def\fps@figure{htbp}
\makeatother

\NewDocumentCommand\citeproctext{}{}

\makeatletter
 \let\@cite@ofmt\@firstofone
 \def\@biblabel#1{}
 \def\@cite#1#2{{#1\if@tempswa , #2\fi}}
\makeatother
\newlength{\cslhangindent}
\newlength{\csllabelwidth}
\newenvironment{CSLReferences}[2] % #1 hanging-indent, #2 entry-spacing
 {\begin{list}{}{%
  \setlength{\itemindent}{0pt}
  \setlength{\leftmargin}{0pt}
  \setlength{\parsep}{0pt}
  \ifodd #1
   \setlength{\leftmargin}{\cslhangindent}
   \setlength{\itemindent}{-1\cslhangindent}
  \fi
  \setlength{\itemsep}{#2\baselineskip}}}
 {\end{list}}
\usepackage{calc}

\newcommand{\CSLLeftMargin}[1]{\parbox[t]{\csllabelwidth}{\strut#1\strut}}
\newcommand{\CSLRightInline}[1]{\parbox[t]{\linewidth - \csllabelwidth}{\strut#1\strut}}

\providecommand{\tightlist}{%
  \setlength{\itemsep}{0pt}\setlength{\parskip}{0pt}}

\pdfoutput=1
\usepackage{caption}
\usepackage{arxiv}
\usepackage{orcidlink}
\usepackage{amsmath}
\usepackage[T1]{fontenc}
\makeatletter
\@ifpackageloaded{caption}{}{\usepackage{caption}}
\AtBeginDocument{%
\ifdefined\contentsname
  \renewcommand*\contentsname{Table of contents}
\else
  \newcommand\contentsname{Table of contents}
\fi
\ifdefined\listfigurename
  \renewcommand*\listfigurename{List of Figures}
\else
  \newcommand\listfigurename{List of Figures}
\fi
\ifdefined\listtablename
  \renewcommand*\listtablename{List of Tables}
\else
  \newcommand\listtablename{List of Tables}
\fi
\ifdefined\figurename
  \renewcommand*\figurename{Figure}
\else
  \newcommand\figurename{Figure}
\fi
\ifdefined\tablename
  \renewcommand*\tablename{Table}
\else
  \newcommand\tablename{Table}
\fi
}
\@ifpackageloaded{float}{}{\usepackage{float}}
\floatstyle{ruled}
\@ifundefined{c@chapter}{\newfloat{codelisting}{h}{lop}}{\newfloat{codelisting}{h}{lop}[chapter]}
\floatname{codelisting}{Listing}

\usepackage{amsthm}
\theoremstyle{definition}
\newtheorem{definition}{Definition}[section]
\theoremstyle{remark}
\AtBeginDocument{}

\makeatother
\makeatletter
\@ifpackageloaded{caption}{}{\usepackage{caption}}
\@ifpackageloaded{subcaption}{}{\usepackage{subcaption}}
\makeatother
\usepackage{bookmark}
\IfFileExists{xurl.sty}{\usepackage{xurl}}{} % add URL line breaks if available
\hypersetup{
  pdftitle={ggtime: A Grammar of Temporal Graphics},
  pdfauthor={Cynthia A. Huang; Mitchell O'Hara-Wild; Rob J. Hyndman; Matthew Kay},
  colorlinks=true,
  linkcolor={blue},
  filecolor={Maroon},
  citecolor={Blue},
  urlcolor={Blue},
  pdfcreator={LaTeX via pandoc}}

\newcommand{\runninghead}{A Preprint }
\title{ggtime: A Grammar of Temporal Graphics}
\def\asep{\\\\\\ } % default: all authors on same column
\author{\textbf{Cynthia A.
Huang}~\orcidlink{0000-0002-9218-987X}\\Econometrics and Business
Statistics\\Monash
University\\Melbourne\\\href{mailto:cynthia.huang@monash.edu}{cynthia.huang@monash.edu}\asep\textbf{Mitchell
O'Hara-Wild}~\orcidlink{0000-0001-6729-7695}\\Econometrics and Business
Statistics\\Monash
University\\Melbourne\\\href{mailto:mitch.ohara-wild@monash.edu}{mitch.ohara-wild@monash.edu}\asep\textbf{Rob
J. Hyndman}~\orcidlink{0000-0002-2140-5352}\\Econometrics and Business
Statistics\\Monash
University\\Melbourne\\\href{mailto:rob.hyndman@monash.edu}{rob.hyndman@monash.edu}\asep\textbf{Matthew
Kay}~\orcidlink{0000-0001-9446-0419}\\Computer Science \& Communication
Studies\\Northwestern
University\\Evanston\\\href{mailto:matthew.kay@u.northwestern.edu}{matthew.kay@u.northwestern.edu}}
\date{}
\begin{document}
\maketitle
\begin{abstract}
Visualizing changes over time is fundamental to learning from the past
and anticipating the future. However, temporal semantics can be
complicated, and existing visualization tools often struggle to
accurately represent these complexities. It is common to use bespoke
plot helper functions designed to produce specific graphics, due to the
absence of flexible general tools that respect temporal semantics. We
address this problem by proposing a grammar of temporal graphics, and an
associated software implementation, `ggtime', that encodes temporal
semantics into a declarative grammar for visualizing temporal data. The
grammar introduces new composable elements that support visualization
across linear, cyclical, quasi-cyclical, and other granularities;
standardization of irregular durations; and alignment of time points
across different granularities and time zones. It is designed for
interoperability with other semantic variables, allowing navigation
across the space of visualizations while preserving temporal semantics.
\end{abstract}

\section{Introduction}\label{sec-intro}

\begin{figure}[t]

\centering{

\pandocbounded{\includegraphics[keepaspectratio]{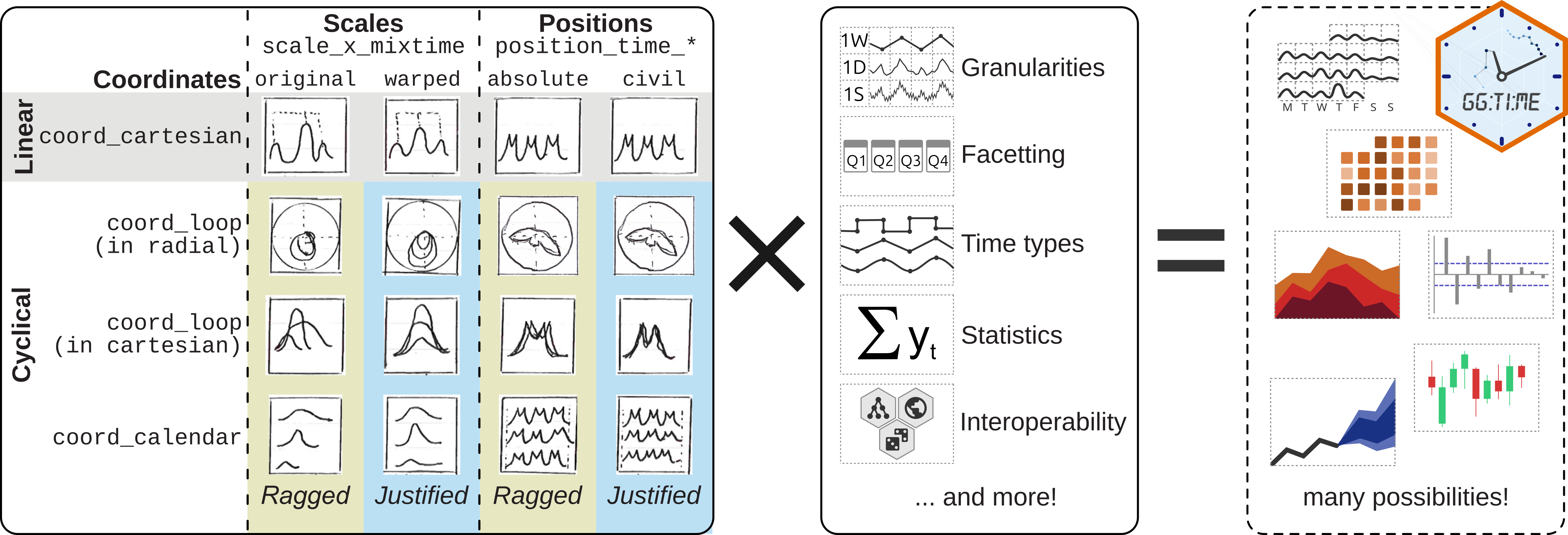}}

}

\caption{\label{fig-teaser}A selection of components in the
grammar-based temporal graphics system {\emph{ggtime}} which can be
combined to produce a variety of visualizations that preserve temporal
semantics. The left box highlights how compositions of Scale, Position,
and Coordinate components highlight different features of repeating
patterns. Each of these components act independently, and can be
combined with other components in {\emph{ggtime}} and the wider
{\emph{ggplot2}} ecosystem to produce a wide variety of temporal
graphics.}

\end{figure}%

Visual representations have long been used to support the analysis and
communication of {\emph{time-oriented data}}. Before the advent of
computing, {\emph{temporal graphics}}, and data visualizations more
broadly were produced by hand. Well known examples include Minard's map
of \emph{Napoleon's march on Russia} {[}28{]} and Nightingale's rose
diagram of \emph{Causes of mortality in the British Army during the
Crimean War} {[}30{]}. In the present day, many programmatic tools,
including graphics libraries, graphical grammars, and chart templates,
have been developed to support visual representations of time-based
information such as models of time (e.g.~calendars, timelines),
empirical observations (e.g.~daily temperature), and statistical models
(e.g.~epidemiological forecasts). There have also been various attempts
to formalize, enumerate, and/or categorize temporal graphics in order to
inform and evaluate such tools. This includes taxonomies of commonly
observed chart types and encoding choices {[}e.g. 44{]}, as well as more
formal characterization of potential visualization design spaces through
frameworks like the \emph{Grammar of Graphics} {[}55{]}.

Wilkinson's \emph{Grammar of Graphics} has inspired a whole category of
programmatic systems and tools for visualization, often referred to as
\emph{graphical grammars}. Graphical grammars allow for declarative
specification and/or construction of graphics via combinations of
grammar components, encompassing both grammar-based interfaces
(i.e.~declarative APIs) and domain specific languages (DSLs)
{[}22,25,37,41,c.f. 57{]}. Notable graphical grammar systems include
{\emph{ggplot2}} {[}54{]}, \emph{plotnine} {[}20{]}, and
\emph{vega-lite} {[}41{]}.

Grammar-based interfaces have been found to be particularly useful for
supporting fluid iteration over different visual encoding choices
{[}37{]}. Furthermore, grammar-based DSLs often integrate formalizations
of validity and good design principles. For example, \emph{ggdist}
{[}18,19{]}, a {\emph{ggplot2}} {[}54{]} extension that implements the
\emph{Probabilistic Grammar of Graphics} {[}37{]}, supports fluid
iteration and exploration over visual representations of statistical
distributions. The ability to quickly and easily iterate over a space of
visualization designs, with the knowledge that certain validity
conditions and/or design principles are maintained, is generally useful
to both data analysts and visualization designers alike. It is
particularly desirable when conducting exploratory time series analysis
because of the complexities of time.

Although our everyday interactions might suggest that time is quite
simple -- an index that increases uniformly, like a simple number line
-- upon closer inspection, various peculiarities, irregularities, and
contradictions become apparent. The quirks of time lead to sometimes
subtle but meaningful challenges in synchronizing temporal data
manipulations with visual encoding choices. There are many different
models of time and time-keeping, each with their own representations of
time and associated temporal operations. These include absolute time
models used in information systems (e.g.~Unix time), calendrical
hierarchies such as the Gregorian calendar, and civil time distortions
like time zones. Even within a single calendar system, relationships
between different temporal units are not always straightforward --
e.g.~in the Gregorian calendar, the number of days in a week is the same
regardless of week, while days in a month depend on the specific month
and year.

Current handling of temporal semantics in graphical grammars is often
ad-hoc and implemented using data preprocessing. Unfortunately, such
approaches can result in the premature loss of important temporal
metadata, leading to inaccurate or misleading visual representations of
time-oriented data. For example, Figure~\ref{fig-wrong-day-order} from
{[}23{]} shows an attempt to plot time-oriented data by day of the week
using the layered grammar of graphics as implemented in {\emph{ggplot2}}
{[}51,54{]}. Notice that the data for each day of the week are plotted
in lexicographical order, rather than calendrical order. This issue
arises because ``day of week'' has been extracted from a date-time as
text, thereby removing important semantic information (the order of days
of the week).

Inspired by prior work that argues visualizations should preserve the
underlying mathematical structure of the data they depict
{[}9,21,24,58{]}, and calls for formal theories of statistical graphics
{[}7,16{]}, we introduce the term {\emph{semantic validity}} to describe
data structures, operations, and visual representations that accurately
reflect the semantics of a particular domain. We formalize three core
design goals for the programmatic creation of {\emph{semantically-valid
graphics}}: \emph{semantic validation}, \emph{fluid navigation}, and
\emph{error friction}. These goals cover a minimal set of feature and
usability requirements for {\emph{semantic visualization systems}}. We
introduce three auxiliary design goals: \emph{generative power},
\emph{interoperability}, and \emph{extensibility} to reflect potential
reasons that many existing semantic visualization systems are
implemented as graphical grammars.

This paper proposes a semantic visualization system for
{\emph{time-oriented data}}\footnote{We adopt the nomenclature of
  ``time-oriented data'' from {[}1{]} to situate this work within the
  InfoVis/HCI literature, though where appropriate we also use the
  statistical nomenclature of ``time series analysis and data.'' We use
  the term ``temporal graphics'' to refer generally to visual
  representations of time and time-based data, encompassing both ``time
  series visualization'' and ``time-oriented visualization.''}, which we
define as data that requires temporal context to accurately interpret or
communicate. The system design consists of formal definitions of
temporal semantics, associated validation conditions, and time-aware
grammar of graphics components that integrate temporal semantics into
Wickham's layered reparameterization {[}51{]} of Wilkinson's
\emph{Grammar of Graphics} {[}55{]}. We implement the system as an
extension to {\emph{ggplot2}} {[}51{]}, the most popular layered grammar
of graphics implementation in R. The R package {\emph{ggtime}} defines
new {\emph{ggplot2}} Geometry, Scale, Position Adjustment, and
Coordinate extension objects. Semantic handling in {\emph{ggtime}} is
also supported by time classes and calendrical operations from the
{\emph{mixtime}} R package {[}31{]}.

We proceed with a review of graphical grammars and related work, as well
as existing visualization systems and tools for time and time-oriented
data. Following this, we define and discuss general design goals for
semantic visualization systems before introducing the key temporal
semantics and validation conditions addressed by {\emph{ggtime}}. We
then explain the conceptual design of our system as a set of time-aware
layered grammar of graphics components and introduce the new
{\emph{ggplot2}} extension objects provided by {\emph{ggtime}}.

Following the description of {\emph{ggtime}}, we discuss the system's
expressivity and generative power, and showcase some of the temporal
visualizations made possible by combining time-aware grammar components.
Selected case studies then illustrate how {\emph{ggtime}} addresses a
number of common pitfalls and mistakes in temporal graphics.

Finally, we conclude with limitations and future directions for this
work, as well as some reflections on interdisciplinary collaboration
between researchers in statistics and information visualization.

\section{Semantic visualization systems}\label{sec-design-goals}

As we will discuss in our review of related work, there is a
well-established need for visualization systems which, in addition to
achieving fluid navigation over visual design spaces, also ensure that
notions of validity are maintained while users iterate over visual
idioms in that space.

Based on this characterization of needs, we define {\emph{semantic
visualization systems}} as graphics systems that encompass the
specification, construction, and rendering of visualizations which
explicitly embed, validate, and preserve specific domain semantics
required to accurately represent and interpret the underlying data or
objects being visualized. We offer the following core design goals for
semantic visualization systems:

\begin{enumerate}
\def\labelenumi{\arabic{enumi}.}
\tightlist
\item
  {\emph{Semantic Validation}}: assess the semantic validity of
  user-specified graphics.
\item
  {\emph{Fluid Navigation}}: minimize cognitive load and effort while
  iterating over valid graphics.
\item
  {\emph{Error Friction}}: increase friction associated with producing
  graphics with invalid domain semantics.
\end{enumerate}

`Semantic Validation' and `Fluid Navigation' can be thought of as
feature and usability requirements respectively, whilst `Error Friction'
is a combination of features and usability. We aim for friction rather
than prevention or avoidance to allow, where appropriate, for flexible
application of validation conditions (e.g.~through warnings rather than
errors). It is not always necessary to strictly enforce validation
conditions on temporal data inputs if correctness would not lead to
perceivable changes in the appearance and/or interpretation of the
resulting visualization. In such cases, strict enforcement can interfere
with the goal of `Fluid Navigation'.

The above goals can generally be achieved by any system that ensures its
data structures, operations, and visual encodings are compatible with
the semantics of the underlying data, and provides a sensible interface
for users to interact with. Such systems are not necessarily restricted
to grammar-based implementations or interfaces. However, many semantic
visualization systems do follow grammar-based design and implementation
approaches. As such, we propose a set of auxiliary design goals for
semantic visualization systems based on potential reasons for this
trend:

\begin{enumerate}
\def\labelenumi{\arabic{enumi}.}
\setcounter{enumi}{3}
\tightlist
\item
  {\emph{Generative Power}}: support the discovery and composition of
  novel visual idioms (with valid semantics) via modularity \&
  composability;
\item
  {\emph{Extensibility}}: support the seamless addition or refinement of
  domain semantics and associated visual encoding options;
\item
  {\emph{Interoperability}}: interface with other semantic visualization
  systems to produce graphics that satisfy validation conditions across
  multiple semantic domains.
\end{enumerate}

`Generative Power' is closely aligned with the expressivity and coverage
of a visualization system, a known benefit of grammar-based interfaces
{[}37{]}. `Extensibility' within a system implies that modifications to
the validation conditions in one part of the visualization system
should, with minimal additional design or engineering considerations, be
congruent with other parts of the visualization system. For instance,
when implementing a new visual design or validation condition,
extensible systems should readily allow interfacing with the temporal
metadata and semantic handling provided by existing data or operation
methods. `Extensibility' also allows for semantic visualization systems
to be built without comprehensive formalization of all semantic aspects
and possible validation conditions for a given semantic domain.
`Interoperability' reflects the reality that many datasets span multiple
semantic domains, and thus support for combinations of domain semantics
is often also required (e.g.~spatio-temporal statistics combines space,
time, and uncertainty). Interoperability can be achieved by implementing
extensions within an existing ecosystem of semantic visualization tools
such as {\emph{ggplot2}}.

These six design goals help to illustrate what makes graphical grammars
such a natural solution framework for semantic visualization systems. As
we will illustrate with this work, `Generative Power', `Extensibility',
and `Interoperability' all arise naturally from the modularity and
composability of grammar-based systems, while `Semantic Validation',
`Fluid Navigation', and `Error Friction' can be achieved through
principled integration of domain semantics and validation conditions
into grammatical rules and components, and careful consideration of
notation design principles such as the \emph{Cognitive Dimensions of
Notation} {[}11{]}.

\section{Related work}\label{sec-related-work}

Our work shares similar motivations and solution approaches with
existing frameworks and systems for working with time-based data and
statistical graphics.

\subsection{Validity in visualization}\label{validity-in-visualization}

There is extensive prior work formalizing and addressing notions of
validity across multiple fields concerned with the visualization of
information and data. Our notion of semantic validity parallels prior
work that holds that mathematical structures in data should be preserved
in some way in the visual form of a visualization. Most generally,
Algebraic Visualization Design (AVD) proposes the idea of visual-data
correspondence, which holds that mathematical structures in underlying
data should correspond closely to mathematical structures in the
perception of visualizations {[}21{]}. Similar notions have been used to
define (for example) \emph{correctness} for probabilistic visualizations
{[}18,37{]}, structure-preserving scale transformations {[}24{]}, and
distance-preserving visual embeddings {[}8,9{]}.

This work also extends existing conceptual foundations for formalizing
silent and visible errors in visualizations arising from incomplete or
inaccurate handling of data semantics more generally. McNutt et al.
{[}26{]} introduce the term {\emph{visualization mirage}} to refer to
``silent \emph{and} significant'' errors, and characterize how such
mirages can arise at different points in the visual analytics process
--- errors that, per {[}36{]}, even experts can make in practice. The
same conceptual model can be applied in the context of statistical time
series analysis and visualization. As discussed in
Section~\ref{sec-usage-slopes}, invalid handling of temporal semantics
can lead to \emph{data-driven} mirages such as misleading slope changes
at daylight saving time (DST) changeovers. Separation of temporal
semantics across multiple variables (e.g.~pre-processing auxiliary
string variables for day-of-week or month to achieve cyclical or
calendrical layouts) can increase the chances of visible encoding
errors.

\subsection{Validity in graphical grammars and the grammar of
(statistical) graphics}\label{sec-related-work-grammars}

In the introductory chapter of {[}55{]}, Wilkinson insists that the
\emph{Grammar of Graphics} should not be seen as a new graphics
scripting language for statistical graphics, but rather a formal theory
of grammatical rules for constructing an organized set of points (the
`graph') mathematically, and then using aesthetic mapping to represent
the graph as a graphic. Wilkinson also stressed that when preparing
statistical graphics, computing statistics before drawing a chart breaks
``the connection between the variables and the graphics that represent
them'', which therefore allows for the possibility of silent errors or
inaccuracies in the resulting graphics. The desire to avoid this type of
error is echoed in many of the visualization tools and systems inspired
by the grammar of graphics (e.g. {[}54{]}, {[}20{]}, {[}41{]}, {[}57{]},
{[}37{]}). However, not all of these tools explicitly support a core
principle of Wilkinson's \emph{Grammar of Graphics}: complete control of
statistics by graphing functions {[}55:106{]}. In particular, including
statistical transformations as part of a declarative specification,
rather than as a data-processing step, is highly varied and often not
included as a defining feature of graphical visualization grammars.
Although {[}57{]} and {[}25{]} both define visualization grammars as
DSLs which specify how to transform and map data into visual marks and
encoding channels, there is no explicit discussion of where in the
specification and rendering process such transformations should be
handled and what transformations should be allowed.

Wilkinson's comments, and the previously articulated design goals, help
to illustrate a subtle but meaningful distinction between different
types of tools and systems under the umbrella of `graphical grammars'
--- grammar-based interfaces that add syntactic sugar on top of existing
tools, domain specific languages (DSLs) that are built with the
modularity and composability of object-oriented design, and `true'
graphical grammars that both define \emph{and} implement formally
defined theories of graphics {[}20,37,41,e.g. 54{]}. Most existing
graphical grammars only support single data tables {[}58{]}, though this
is not a restriction posed conceptually by the \emph{Grammar of
Graphics}, and some recent work attempts to extend grammar-based
visualization approaches beyond single data tables. Wu et al.'s
\emph{Formalization and Library for Database Visualization} {[}58{]}
defines the notion of \emph{faithfulness} to refer to visualizations
that preserve underlying database constraints, and motivates their work
by the desire to treat databases as complete inputs to visualization
systems. In effect, they contribute a visualization model and
implementation for semantically valid visualization of databases, with
validation constraints based on an existing formalization: the
relational data model. The authors' assessment and criticisms of current
graphical grammars also exactly echo Wilkinson's warnings against
breaking the connection between data and visualization, as well as prior
calls for theories of graphics {[}7{]}. There have also been attempts to
visualize graph-based semantic objects such as Causal-Loop Diagrams
(representations of graph-based conceptual system models) {[}5{]} and
Crossmaps (representations of ex-post data harmonization operations)
{[}15{]}.

\subsection{Temporal data analysis and
visualization}\label{temporal-data-analysis-and-visualization}

\emph{TimeViz Browser 2.0} {[}44{]} is a recent visual survey of 161
techniques for time-based visualization, drawing primarily from
Information Visualization, Visual Analytics, and related fields. As
explained in {[}1{]}, the survey is organized according to a simplified
categorization schema designed to guide the selection of visual
encodings and chart types, rather than accurately reflect the full
complexity of temporal graphics. The schema prioritizes the \emph{what}
(time and data) and \emph{how} (visual representation) aspects of the
visualization problem, and purposely leaves the \emph{why} (user tasks)
as a separate consideration. By contrast, the motivation for our work
arose in the context of exploratory time series analysis and, as such,
must follow the well established need for a systematic and principled
approach to handling statistical and temporal semantics that cannot be
separated from the \emph{why} {[}7,16,55{]}.

In this regard, our work is most similar in spirit to Wills's {[}56{]}
attempts to provide \emph{prescriptive} advice on designing graphical
representations for statistical data (with a time dimension) under the
\emph{Grammar of Graphics} {[}55{]} framework. Wills details various
ways temporal semantics might be incorporated into the grammar of
graphics, including using time as 1-D (e.g.~time as the horizontal axis)
and 2-D coordinate spaces (e.g.~via faceting), mapping time to
aesthetics (e.g.~years to color), and distortions of time
(e.g.~normalization of monthly data to account for different numbers of
days). Although developed independently, the design of {\emph{ggtime}}
and the underlying grammar of temporal graphics adheres to many of the
suggestions and principles articulated in {[}56{]}. However, in contrast
with our work, {[}56{]} implements these ideas using \emph{VizML}
{[}13{]}, and as such does not consider how to parameterize temporal
semantics with layered grammar of graphics components or the development
challenges of integrating them into an existing ecosystem like
{\emph{ggplot2}}. We also note that {[}56{]} does not address the
handling and visual representation of multiple timescales (e.g.~daylight
saving time zones) or multiple granularities (e.g.~daily and monthly
data) in a single temporal graphic, while proper handling of multiple
granularities is fundamental to {\emph{ggtime}}.

Our work also shares some common goals and design principles with
software for time-based data analysis and forecasting, as well as
specialized systems and libraries for working with time-based data. On
the analysis side, this work builds upon support for time series
analysis in the R programming language; notably the \emph{tsibble}
{[}48,50{]}, \emph{feasts} {[}33{]}, and \emph{fable} {[}32{]} packages,
which provide data structures and methods for working with time series
data following ``tidy'' workflows {[}53{]}. These packages provide
functions for producing commonly used visualizations for time series
data, such as seasonal plots, autocorrelation plots, and forecast plots.
However, these functions are limited to known and commonly used
visualization techniques, and can be difficult to customize or combine
with other {\emph{ggplot2}} objects. We also build upon existing
attempts to extend {\emph{ggplot2}} to support calendrical layouts
{[}49{]} and mixed granularity temporal graphics {[}12{]}.

The design considerations addressed in our work also overlap with those
of general-purpose and domain specific systems for working with
time-based data. Examples of time-specific systems are numerous and
covered extensively in {[}1{]} and {[}44{]}. Most such systems are built
for specific application contexts (e.g.~epidemiological tracing, demand
forecasting, or climate modeling) and have less complete support for
temporal semantics than a more general grammar. Some prior work attempts
to fully address the complexities of time-based data within programmatic
tools for time-based data visualization and analysis. Although primarily
focused on interactive temporal graphics, which are outside the direct
scope of this work, \emph{TimeBench} {[}40{]} and \emph{time-i-gram}
{[}43{]} are notable existing attempts at semantic visualization systems
for time-oriented data. Both ensure temporal semantics are respected
throughout the data, operation, visualization, and interaction to
support significantly more complex temporal semantics than standard data
visualization libraries. However, as noted in the motivation for
\emph{time-i-gram}, \emph{TimeBench} is heavily tied to a Java
implementation, with attendant issues with web integration. In contrast,
\emph{time-i-gram} is proposed as a ``declarative time-based grammar''
for multiscale interactive time visualizations consisting of six
consecutive components: \emph{data types and variables}, \emph{visual
marks}, \emph{visual channels}, \emph{scales}, \emph{layouts}, and
\emph{interaction}. Although this approach shares some similarities with
ours, it has a slightly more abbreviated parameterization compared to
the \emph{layered grammar of graphics} and lacks the built-in
interoperability of the {\emph{ggplot2}} extension ecosystem.

\section{Formal notions of time and time-oriented
data}\label{sec-semantics}

In order to reason about how characteristics of time can and should be
represented visually, and what data structures are needed to correctly
handle time-oriented data, we briefly introduce some foundational
concepts, drawing most directly on {[}1{]} and {[}12{]}. We also note
that there is a long history of prior work across mathematics, computer
science, and associated subfields formalizing these concepts. This
includes attempts to comprehensively formalize time using applied
mathematical logic {[}e.g. 2,35{]}, integrating temporal reasoning into
databases {[}e.g. 3{]}, developing programmatic support for all types of
calendars and calendrical calculations {[}39{]}, and even standardizing
nomenclature for temporal concepts {[}17{]}.

\subsection{Chronons and granules}\label{chronons-and-granules}

We start with discrete time, where the smallest indivisible unit of time
is a {\emph{chronon}} (e.g.~a second, day, month, or year). Without loss
of generality, we can represent the chronons by the integers, with some
arbitrary origin (e.g.~Unix {\emph{epoch}}). In order to properly
characterize the semantic properties of time, we need to formally define
{\emph{granularities}} as mappings that partition chronons into subsets.

\begin{figure}

\centering{

\pandocbounded{\includegraphics[keepaspectratio]{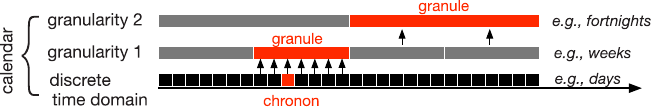}}

}

\caption{\label{fig-aigner-chronon-granules}Illustrative example of how
subsets of chronons form granularities. In this graphic, 1 day chronons
are mapped to weekly (7 days) and fortnightly (2 week) granularities.
This figure is sourced from {[}1{]}.}

\end{figure}%

\subsection{Granularities}\label{granularities}

\begin{definition}[]\protect\hypertarget{def-timedomain}{}\label{def-timedomain}

A {\emph{\emph{time domain}}} is a pair \((T;\leq)\), where \(T\) is a
non-empty set of chronons and \(\leq\) is a total order on \(T\). Let
\({Z} = \{z \in \mathbb{Z}\}\) be the \emph{index set} of integers that
uniquely maps all chronons in \(T\) to integers.

\end{definition}

\begin{definition}[]\protect\hypertarget{def-granularity}{}\label{def-granularity}

A {\emph{\emph{granularity}}} is a mapping \(G\) from the integers (the
\emph{index set}) to subsets of the time domain. Each non-empty subset
\(G(i)\) is a {\emph{\emph{granule}}}.

\end{definition}

The illustration in Figure~\ref{fig-aigner-chronon-granules} from
{[}1{]} demonstrates the relationship between the concepts of chronons,
granularities, and granules.

\begin{definition}[]\protect\hypertarget{def-linear-granularity}{}\label{def-linear-granularity}

A {\emph{\emph{linear granularity}}} is a granularity that satisfies the
following two properties for all \(i,j,k \in Z\):

\begin{enumerate}
\def\labelenumi{(\arabic{enumi})}
\tightlist
\item
  if \(i < j\), \(G(i)\ne G(j)\), and \(G(i)\) and \(G(j)\) are
  non-empty, then each element of \(G(i)\) is less than all elements of
  \(G(j)\); and
\item
  if \(i < k < j\), \(G(i)\ne G(j)\), and \(G(i)\) and \(G(j)\) are
  non-empty, then \(G(k)\) is non-empty.
\end{enumerate}

\end{definition}

This definition implies that the granules in a linear granularity must
be {\emph{\emph{non-overlapping}}}, {\emph{\emph{contiguous}}},
{\emph{\emph{totally ordered}}}, and {\emph{\emph{non-repeating}}}. For
example, if the chronons in a given time domain are hours, it is
possible to define the linear granularities day, week, month, and year.

\begin{definition}[]\protect\hypertarget{def-circular-granularity}{}\label{def-circular-granularity}

A {\emph{\emph{circular granularity}}} is a granularity such that
\(G(i) = G(j)\) if and only if \(i \equiv j \mod p\). Each non-empty
subset \(G(i)\) is called a {\emph{circular granule}}.

\end{definition}

Examples from the Gregorian calendar include day-of-week and
month-of-year. It is also possible to have \emph{quasi-circular}
granularities such as day-of-month or day-of-year, where the periodicity
\(p\) can vary with \(i\). In contrast to linear granularities, circular
and quasi-circular granularities are {\emph{\emph{repeating}}} and
{\emph{\emph{cyclically ordered}}} (modulo \(p\)), rather than
{\emph{\emph{non-repeating}}} and {\emph{\emph{totally ordered}}}.

\subsection{Calendars}\label{calendars}

Using the above definitions, {\emph{calendars}} can be defined as a
system of multiple granularities in lattice structures that specify
relationships between granularities (e.g.~seconds-to-minutes). In other
words, calendars are mappings between a time domain and human-meaningful
time units. The most widely used calendar is the Gregorian calendar, but
other common calendars include the Julian, Islamic, Hebrew, and Chinese
calendars. Some calendars are based on astronomical phenomena, such as
lunar or solar cycles, while others are based on religious or cultural
traditions. Most involve a combination of these factors.

Some of these calendars may include granularities that are neither
linear nor circular. Such granularities may be non-repeating and
unordered. For example, public holidays form an unordered binary
granularity (holiday vs.~non-holiday) that can vary from year to year
(subject to government decree), and are therefore non-repeating. Another
important type of calendar is a {\emph{censored calendar}}, which
systematically omits periods of time from a standard calendar with a set
of rules. Examples of censored calendars include stock market trading
days (omitting weekends and public holidays) or business hours (9am --
5pm weekdays).

\subsection{Discrete and continuous time
models}\label{discrete-and-continuous-time-models}

Continuous time models allow for arbitrarily small subdivisions of time,
in contrast to discrete time models which quantize continuous time by
chronons. By substituting chronons for continuous time intervals, we can
apply the definitions of granularities to continuous time models as
well.

\subsection{Computational models of time}\label{sec-semantics-comp}

The goal of modeling time in information systems is not to perfectly
imitate time, but to accurately represent the characteristics of time
most relevant to the goals of the system {[}1{]}. Most information
systems tend to use discrete time domains (e.g.~Unix time is defined
using the number of non-leap seconds since the Unix {\emph{epoch}}).
Some models aim to represent and track a universal notion of time, often
referred to as {\emph{absolute time}} (e.g.~POSIX time and International
Atomic Time (TAI)), whilst others, referred to as {\emph{civil time}}
models, aim to reflect and capture socially constructed systems of
timekeeping such as the Gregorian calendar and time zones (e.g.~the ISO
8601 standard). We limit the scope of this work to representations and
analysis of time-oriented data in absolute and civil time. This scope
excludes models of {\emph{relative time}}, which describe time relative
to another time point (e.g.~5 minutes from now), as well as time
distortions resulting from motion (\emph{special relativity}) or
gravitational fields (\emph{general relativity}).

\section{Temporal validity}\label{sec-validity}

In this section, we define a number of properties or conditions relevant
for semantically valid visualization of time and time-oriented data.

\begin{figure}

\begin{minipage}{0.50\linewidth}

\centering{

\includegraphics[width=0.9\linewidth,height=\textheight,keepaspectratio]{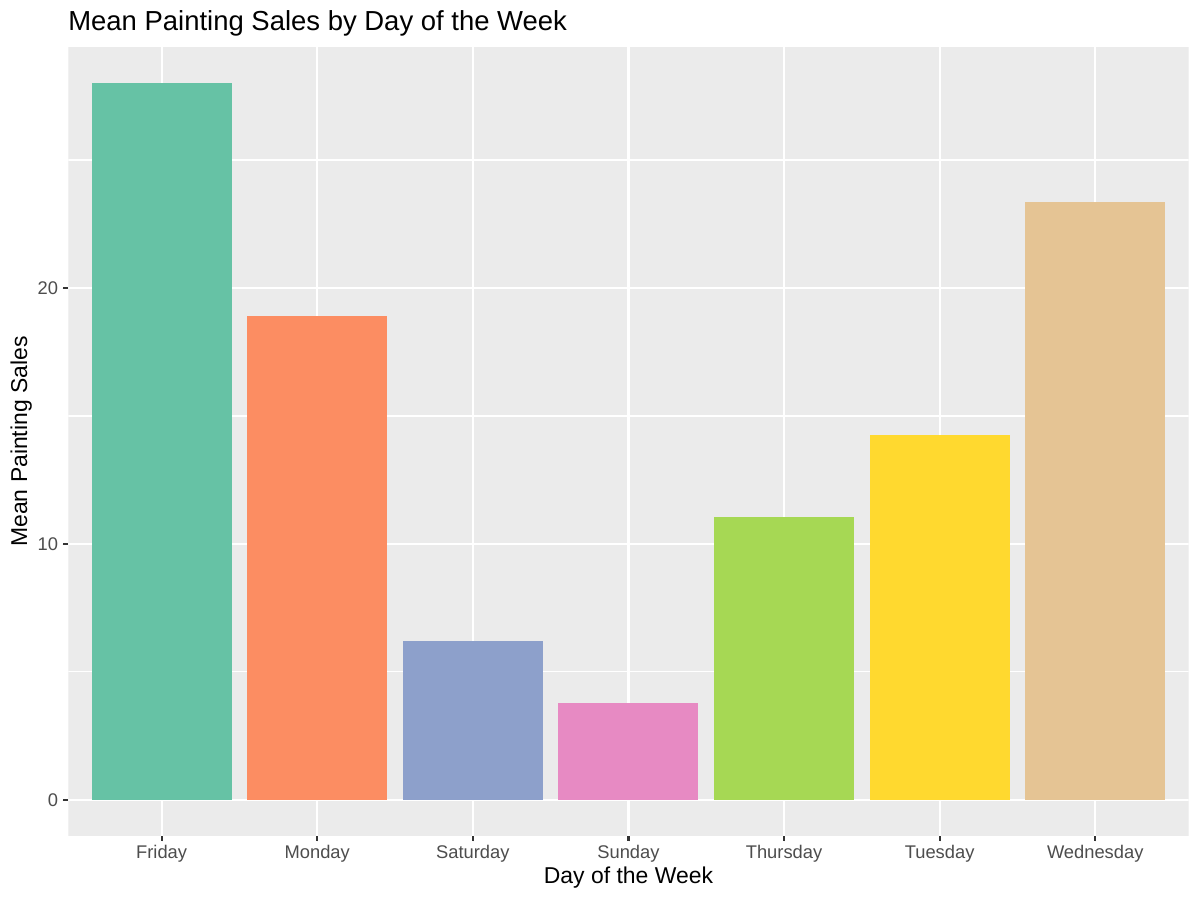}

}

\subcaption{\label{fig-wrong-day-order}Incorrect lexicographical
ordering of days of the week from lossy discretization of time series}

\end{minipage}%
\begin{minipage}{0.50\linewidth}

\centering{

\includegraphics[width=0.9\linewidth,height=\textheight,keepaspectratio]{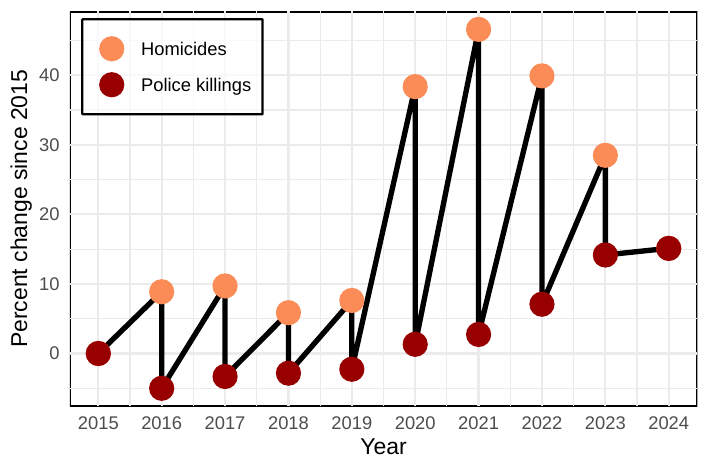}

}

\subcaption{\label{fig-sawtooth-groups}Two distinct time series drawn as
one series due to a missing group produces a `saw tooth' appearance as
observations are not unique}

\end{minipage}%
\newline
\begin{minipage}{0.50\linewidth}

\centering{

\includegraphics[width=0.9\linewidth,height=\textheight,keepaspectratio]{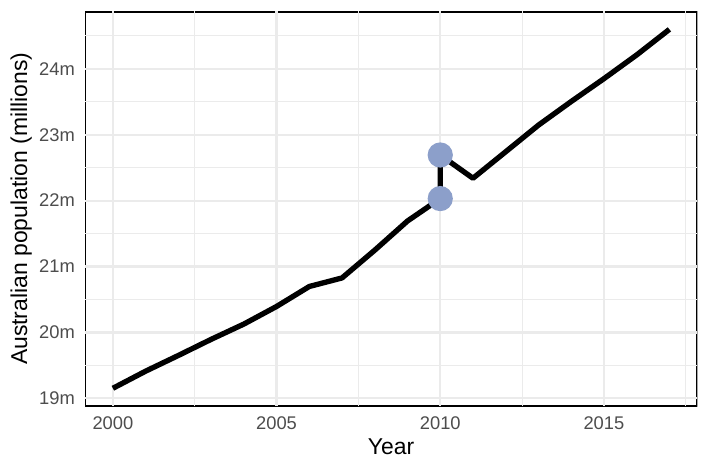}

}

\subcaption{\label{fig-sawtooth-duplicate}A data error where multiple
values for the same time point also introduces the `saw tooth' shape}

\end{minipage}%
\begin{minipage}{0.50\linewidth}

\centering{

\includegraphics[width=0.9\linewidth,height=\textheight,keepaspectratio]{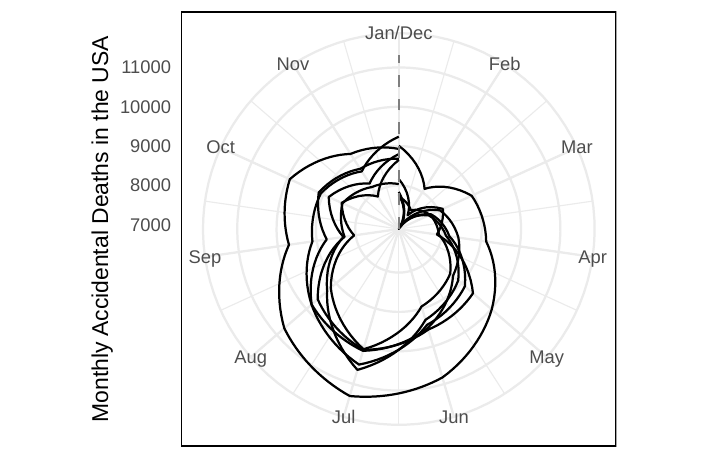}

}

\subcaption{\label{fig-polar-months-wrong}Conversion to circular
granularities is lossy, and when stored numerically without temporal
semantics the first and last seasons are drawn overlapping.}

\end{minipage}%

\caption{\label{fig-gallery-of-bad-plots}A collection of temporal
graphics with mistakes or ambiguities arising from violations of
validity conditions.}

\end{figure}%

Visual encodings (such as axes, positions, and shapes in plots) must
accurately reflect relevant properties of time to avoid creating
inaccurate or incomplete visual representations of time and/or time
series data. For example, Figure~\ref{fig-wrong-day-order} from {[}23{]}
shows an attempt to plot time-oriented data by day of the week using
ggplot2, but the days of the week are plotted in lexicographical order
rather than cyclical time order. A similar visual inaccuracy is shown in
Figure~\ref{fig-polar-months-wrong}, where the months of December and
January share the same position. These types of errors can be avoided by
preserving temporal semantics throughout temporal operations, returning
a circular granularity rather than text or numbers when extracting
circular granularities (e.g.~day of week) from a linear granularity.

We use the following previously defined properties and conditions of
time domains to characterize validity in temporal graphics:

\begin{itemize}
\tightlist
\item
  {\emph{Continuous}}: time which is infinitely divisible into more
  precise measurements (in continuous time models)
\item
  {\emph{Contiguous}}: time which flows without gaps (in discrete time
  models)
\item
  {\emph{Ordered}}: time must flow from past to present (linear) or
  cyclically (circular)
\item
  {\emph{Complete}}: represents all chronons within a given granularity
\end{itemize}

Mistakes and/or ambiguities in temporal graphics can also arise from
issues in the quality, interpretation, or transformations of the
underlying time series data being visualized. We use the following
properties of time series data to discuss such issues:

\begin{itemize}
\tightlist
\item
  {\emph{Temporally Unique}}: there is at most one observation per
  chronon.
\item
  {\emph{Temporally Determinate}}: each observation is associated with a
  single chronon.
\end{itemize}

A common visual indication of mistakes in line charts of time series
data (e.g.~macroeconomic indicators, sales figures, mortality counts)
are jagged `saw-tooth' artifacts. {\emph{Saw-toothing}} generally arises
from some violation of temporal uniqueness. For example,
Figure~\ref{fig-sawtooth-groups} shows saw-toothing from failing to
separate different time series. Violations of uniqueness can also occur
due to data quality issues. For example, consider visualizing a time
series containing measurements with two observations for the same
timestamp (e.g.~due to accidental row duplication). The two observations
will be connected by a vertical jump as shown in
Figure~\ref{fig-sawtooth-duplicate}. This type of vertical jump can also
be caused by daylight saving time changeovers, as discussed in
Section~\ref{sec-usage-slopes}.

{\emph{Temporal indeterminacy}} refers to imprecision or uncertainty in
a statement with respect to time {[}17{]}. For example, statements of
time within an interval are indeterminate (e.g.~an event happens between
3--5pm). Operations involving mappings between time models of different
precision, such as conversions between coarser chronons (e.g.~months)
into finer chronons (e.g.~days) or to continuous time models, can also
give rise to temporal indeterminacy. As we discuss in
Section~\ref{sec-usage-nyt}, it is generally not straightforward to
visualize temporally indeterminate data, particularly with static
graphics.

\section{A graphical grammar system for temporal
graphics}\label{sec-system-description}

As explained in Section~\ref{sec-related-work-grammars}, the modularity
and composability of graphical grammars enable principled integration of
semantics into the visualization process. We design and implement our
system as an extension of the \emph{Layered Grammar of Graphics} and its
R implementation, {\emph{ggplot2}} {[}51,54{]}. The use of a
well-established and mature graphical grammar system allows us to focus
on the conceptual integration of temporal design aspects and validation
conditions into appropriate layered grammar of graphics components.

To characterize the scope of our temporal visualization system, we
follow Aigner et al. {[}1{]} in defining four fundamental design aspects
for time-oriented data visualization: \emph{scale}, \emph{scope},
\emph{arrangement}, and \emph{viewpoint}. Scale addresses whether time
is depicted as continuous, discrete, or ordinal, influencing the
granularity at which moments or intervals are represented. Scope
concerns whether the temporal domain is point-based or interval-based, a
distinction critical to expressing durations. Arrangement refers to how
time is structured visually, be it linearly (i.e.~past-to-future) or
cyclically (such as seasons or recurring events). Finally, viewpoint
addresses the choice of perspectives being presented. This includes
restrictions to totally ordered sequences (e.g.~one event after another)
and partially ordered (overlapping events). Viewpoint also addresses
depictions of temporal indeterminacy such as branching (i.e.~parallel
timelines), or multiple perspectives (e.g.~inconsistent witness accounts
for the time of an event).

Our system aims to support a wide range of these temporal design aspects
within the grammar of graphics framework. We cover both discrete and
continuous temporal scales and offer some support for ordinal
representations of time --- though the creation of highly idiosyncratic
timelines is generally better supported by interactive tools
(e.g.~\emph{TimeSplines} {[}34{]}). The current implementation supports
only point-based time domains, but handling of interval-based
visualizations is planned. Both linear and cyclical arrangements are
supported by {\emph{ggtime}}, enabling common use cases ranging from
simple timelines to calendar or circular plots. For viewpoint,
{\emph{ggtime}} reliably supports totally ordered time and can handle
some partially ordered cases (such as representations involving civil
DST shifts as discussed in Section~\ref{sec-usage-slopes}). However,
more complex branching or multi-perspective scenarios are best handled
through the combination of graph theory and temporal semantics, and as
such are beyond the direct scope of {\emph{ggtime}}.

\begin{figure}

\centering{

\pandocbounded{\includegraphics[keepaspectratio]{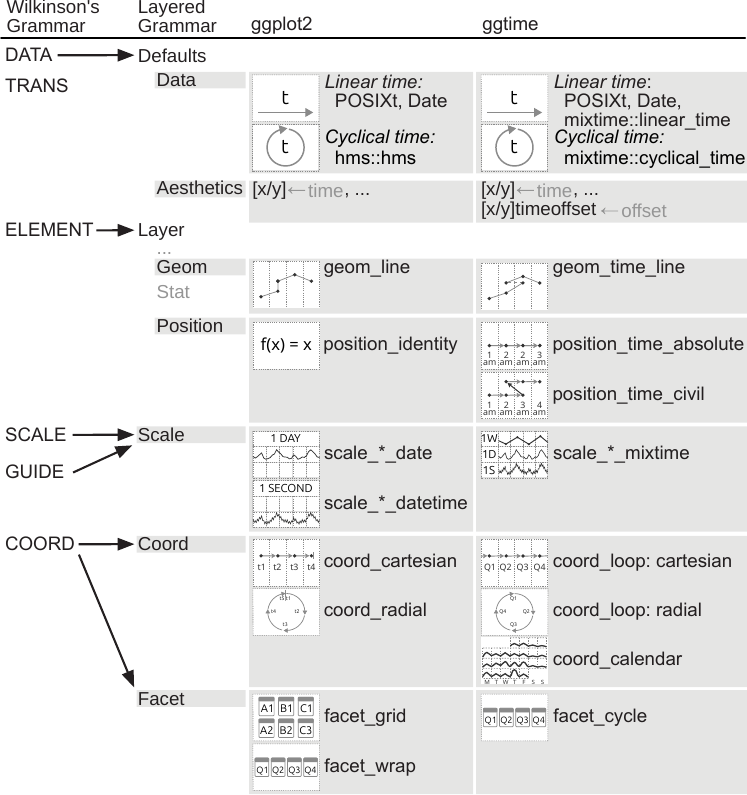}}

}

\caption{\label{fig-ggtime-component-icons}A visual summary of relevant
grammatical elements for temporal visualization provided by ggplot2 and
extended by ggtime. The elements are organized into grammar components
as described in the layered grammar of graphics {[}51{]}.}

\end{figure}

In this section we introduce the operationalized design and
implementation of our proposed temporal visualization system as
extension components within {\emph{ggplot2}}.
Figure~\ref{fig-ggtime-component-icons} summarizes the existing
{\emph{ggplot2}} components for temporal visualization and the
extensions provided by {\emph{ggtime}}.

\subsection{Data}\label{sec-grammar-data}

The data component provides the observations for the graphic, and the
data type encodes the semantics for how it should be visualized.
Time-oriented data can encode many different temporal semantics
described in Section~\ref{sec-semantics}, including the time model
(discrete or continuous), granularity (linear or circular), and calendar
system (e.g.~Gregorian). The data also brings with it associated methods
for validating and transforming time, which is needed within other
grammar components.

We address the need to preserve temporal metadata in {\emph{ggtime}} by
storing time with flexible temporal {\emph{mixtime}} vectors and
temporally self-validating \texttt{tsibble} data frames {[}48{]}. These
data structures encode all four design aspects defined by {[}1{]}.
{\emph{mixtime}} implements an extensible system for storing time in
vectors, allowing time observed at different granularities, time zones,
and calendar systems to coexist within the same variable. New linear and
cyclical {\emph{arrangements}} of time can be created with
\texttt{time\_linear()} and \texttt{time\_cyclical()}, each of which
accepts discrete (integer type) or continuous (double type) time
{\emph{scales}}. The {\emph{scope}} of the time domain is point-based,
but interval-based time can be created with the \emph{ivs} package
{[}46{]}. \emph{tsibble} is a rectangular data structure that accepts
mixtime vectors as time indices and a set of key variables for
identifying multiple time series within the same data frame. It is
self-validating, ensuring that time indices are unique for each key, and
provides tools to check and repair completeness. The identifying keys of
a \emph{tsibble} enable more complex {\emph{viewpoints}} of time, where
key variables can identify different perspectives (e.g.~witness
accounts) or branches (e.g.~parallel timelines).

\subsection{Aesthetics}\label{sec-grammar-aesthetics}

The aesthetics map variables from the data to {\emph{visual channels}}
in the graphic. Existing aesthetics in {\emph{ggplot2}} provide suitable
mappings for different {\emph{scopes}} of time-oriented data, including
point-based (e.g.~\texttt{x}, \texttt{y}, \texttt{color},
\texttt{shape}) and interval-based (e.g.~\texttt{xmin}, \texttt{xmax},
\texttt{ymin}, \texttt{ymax}) aesthetics.

The positional aesthetics are extended by {\emph{ggtime}} with
\texttt{xtimeoffset} and \texttt{ytimeoffset}, which encode relative
time adjustments to a reference time mapped to \texttt{x} and \texttt{y}
(e.g.~\texttt{+1} hour, \texttt{-1} hour, etc.). These positional
aesthetics are primarily used to disambiguate civil time ordering during
transitions between time zones (e.g.~daylight savings time shifts), but
they can also encode the re-calibration of time from clock
synchronization. Mapping absolute time in a reference timezone
(e.g.~GMT) to \texttt{x}/\texttt{y} and the relative offset for the
local timezone to \texttt{xtimeoffset}/\texttt{ytimeoffset} allows
geometries to preserve the {\emph{temporally unique}} and
{\emph{ordered}} properties of time; this can be done directly by a user
or automatically using \texttt{position\_time\_civil()}
(Section~\ref{sec-grammar-position}).

\subsection{Geometries}\label{sec-grammar-geometries}

Geometries are the {\emph{visual marks}} used to display observations in
the graphic (e.g.~points, lines, bars, areas). Existing {\emph{ggplot2}}
geometries can be used to represent different {\emph{scopes}} of
time-oriented data, including point-based (e.g.~\texttt{geom\_point()},
\texttt{geom\_area()}) and interval-based (e.g.~\texttt{geom\_rect()},
\texttt{geom\_ribbon()}). While line geometries provided by
{\emph{ggplot2}} can represent sequences of observations, they lack
necessary capabilities for safely encoding temporal validity. For
example, the \texttt{geom\_line()} geometry sorts the data in (time)
order when mapping time to the \texttt{x} aesthetic, but incorrect
graphics can arise when the data is not {\emph{temporally unique}} or
{\emph{contiguous}}, including saw-toothing (Section~\ref{sec-validity})
and connected lines between discontiguous observations.

The {\emph{ggtime}} package implements \texttt{geom\_time\_line()}, a
time-aware extension of \texttt{geom\_line()} which produces accurate
slopes in the presence of discontiguities, duplications, and misordering
in time. For example, \texttt{geom\_time\_line()} uses changes in
offsets mapped to \texttt{xtimeoffset} and \texttt{ytimeoffset} to
identify discontiguities, which it renders as a dashed line parallel to
the time axis. Section~\ref{sec-usage-slopes} demonstrates how this
maintains accurate slopes when visualizing civil time.

\subsection{Statistics}\label{sec-grammar-statistics}

Statistics are {\emph{transformations}} of data variables before they
are mapped to aesthetics. Graphical statistics typically involve
aggregation or summarization (e.g.~binning, counting, or smoothing).
Temporal aggregation typically involves grouping observations into
coarser granularities (e.g.~aggregating daily observations into monthly
averages) and is commonly used to reduce the visual noise arising from
fine-grained patterns.

The {\emph{ggtime}} package currently does not implement any
time-specific visual statistics, since statistical summaries are already
well supported in data pre-processing of {\emph{mixtime}} vectors using
\texttt{tsibble} and \texttt{dplyr}. However, commonly used statistical
summaries over time are considered in
Section~\ref{sec-future-grammar-components} as future directions for
{\emph{ggtime}}, including \texttt{stat\_ohlc()} for open-high-low-close
summaries used in financial time series visualization.

\subsection{Position}\label{sec-grammar-position}

Position adjustments modify the placement of data values after all other
mappings and transformations have been applied. {\emph{ggtime}} provides
position adjustments to present data in either civil time
(\texttt{position\_time\_civil()}) or absolute time
(\texttt{position\_time\_absolute()}), which can be used with any
geometry from ggplot2 and its ecosystem of extensions. These position
functions use timezone information attached to {\emph{mixtime}} and
\texttt{POSIXct}\footnote{\texttt{POSIXct} is a built-in \emph{R} data
  type representing time in seconds since the Unix epoch.} time classes
to apply the relative offset from the reference timezone. Because
{\emph{mixtime}} stores timezone information with all temporal
granularities,\footnote{Unlike \texttt{Date} and other
  non-\texttt{POSIXct} time classes in \emph{R} and other programming
  languages.} the absolute timing of dates can be more accurately
positioned across multiple time zones.

\subsection{Scale}\label{sec-grammar-scales}

Scales map data values to aesthetic values and are tightly coupled with
the guides of the graphic for axis ticks and labels. The scale component
is primarily responsible for mapping different time {\emph{scales}}
(continuous, discrete, ordinal) to continuous numeric positions along
positional axes (\texttt{x}, \texttt{y}), and other aesthetics
(e.g.~\texttt{color}, \texttt{fill}). This process is straightforward
for single granularity time data, since the internal representation of
time is already a numeric value of the number of time units
(e.g.~seconds for \texttt{POSIXct} and days for \texttt{Date}) from an
origin (e.g.~Unix {\emph{epoch}}).

The generality of {\emph{mixtime}} vectors complicates this mapping,
since time measured at different granularities, in different time zones,
and even in different calendar systems can coexist within the same
vector. As such, the scale component needs to handle the conversion of
all observations to a common timescale before they can be mapped to
aesthetic values. The \texttt{scale\_*\_mixtime()} functions in
{\emph{ggtime}} implement this by first converting all times from
semantically discrete to continuous time models, and then using
functionality from {\emph{mixtime}} to convert all observations to a
common temporal granularity. The common granularity is automatically
identified using the greatest lower bound among the chronons {[}4{]},
which is the coarsest granularity that all observations can be mapped to
(e.g.~days are common to weeks and months). The process of converting
from discrete to continuous time models is {\emph{temporally
indeterminate}}, since it involves selecting a specific moment within
the coarser granularity to use for the mapping (e.g.~which day in
January should represent the whole month of January?). The default
behavior is to use the middle of the granularity, which can be adjusted
with the \texttt{align\_mixed} argument that accepts any value 0-1 (0 =
start, 0.5 = middle, 1 = end). Section~\ref{sec-usage-nyt} demonstrates
how different alignment choices can affect the interpretation of
graphics with mixed-granularity time data.

The scale for mixtime objects also supports the normalization of time
that spans intervals of varying lengths. This process is generally known
as both {\emph{time warping}} and {\emph{curve registration}} {[}38{]},
and is useful for aligning patterns that repeat with irregular
periodicity. Specific time points between which time is normalized to be
within 0-1 can be specified with the \texttt{warps} argument of
\texttt{scale\_*\_mixtime()}. Suitable time points are typically known
{\emph{landmarks}} in the cyclical pattern being visualized, such as
peaks or troughs. Calendrical granularities can also be used with the
\texttt{time\_warps} argument to normalize the length of irregular
granularities (e.g.~days in months). Time warping is particularly useful
in combination with circular coordinate systems, as demonstrated in
Section~\ref{sec-design-space}.

\subsection{Coordinate}\label{sec-grammar-coordinates}

\begin{figure}

\centering{

\pandocbounded{\includegraphics[keepaspectratio]{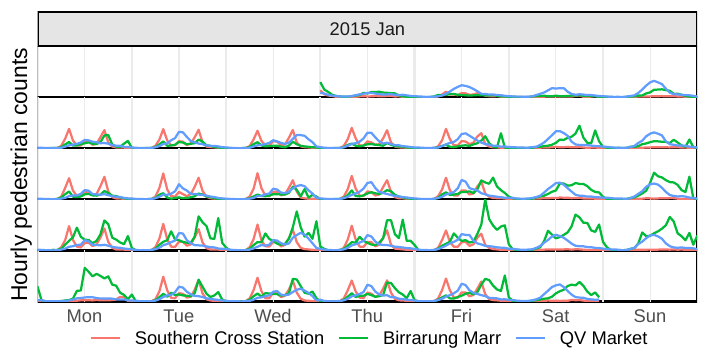}}

}

\caption{\label{fig-good-calendar}A calendar coordinate system separates
granules into individual rows. This figure shows three time series in a
calendar arrangement, where each row contains a week of hourly
observations.}

\end{figure}%

Coordinates define the coordinate system used to transform positions
onto a static 2D graphic. Alternatives to Cartesian coordinate systems
are often needed when a semantic variable contains information requiring
a projection to be shown in a two-dimensional space. For example,
latitudes and longitudes in spatial data need to be projected onto a
flat surface for visualization with one of many map projections. The
Cartesian coordinate system (\texttt{ggplot2::coord\_cartesian()}) is
suitable for visualizing linear {\emph{arrangements}} of time, however,
it is not well suited for visualizing repeating patterns in circular
granularities since the Cartesian coordinate system is non-repeating.
The polar or radial coordinate systems (\texttt{ggplot2::coord\_polar()}
/ \texttt{ggplot2::coord\_radial()}) are more suitable for visualizing
circular arrangements of time, since the circular nature of the
coordinate system better reflects the cyclical semantics of the data.
However, the inferiority of polar and radial coordinates for visual
perception and accuracy of interpretation is well documented {[}47{]},
and so Cartesian coordinates are generally preferred even when
visualizing circular granularities. Visualizing seasonal patterns with
circular granularities in Cartesian coordinates requires cumbersome data
pre-processing as shown in Section~\ref{sec-usage-circular} and is prone
to errors (see Figure~\ref{fig-wrong-day-order}).

The {\emph{ggtime}} package provides two coordinate systems for
visualizing cyclical arrangements of time: \texttt{coord\_loop()} and
\texttt{coord\_calendar()}. The looped coordinate system
(\texttt{coord\_loop()}) loops Cartesian (default) or polar/radial
coordinates around specific time points, allowing linear time
granularities to be projected into a circular arrangement. Observations
between each loop are superimposed, allowing the shape of repeating
patterns in time to be easily compared. This allows seasonal and
cyclical patterns to be visualized without needing to first convert time
to circular granularities, which has the added benefit of preserving the
otherwise lost {\emph{contiguity}} between seasons. This allows lines
connecting across seasons to be drawn (e.g.~connecting December to
January) with appropriate temporal spacing, preventing common violations
of semantic validity like those shown in
Figure~\ref{fig-wrong-day-order} and
Figure~\ref{fig-polar-months-wrong}. Specific time points to loop around
can be specified with the \texttt{loops} argument, and common
calendrical granularities can be used with the \texttt{time\_loops}
argument (e.g.~\texttt{"1\ year"}) --- thus unlike
\texttt{ggplot2::coord\_radial()}, irregular periods are supported (see
Section~\ref{sec-design-space-ragged}). A variant of the looped
coordinate system is \texttt{coord\_calendar()}, which separates each
loop into its own row and/or column, creating a dense layout that
resembles a calendar. Reading the calendar linearly (left to right, top
to bottom) provides more visual space to see fine-grained details such
as anomalies and special events/holidays, while reading the calendar
cyclically (top to bottom, left to right) allows the shape of each
season to be compared.

\subsection{Facet}\label{sec-grammar-facets}

Facets create multiple subplots based on one or more
discrete/categorical variables. They are commonly used when visualizing
time series in order to separately plot related time series of different
scales. Faceting by circular granularities of time can also be useful in
showing repeating patterns, particularly for observing changes in the
pattern over time. Existing faceting functions in {\emph{ggplot2}}
(\texttt{facet\_wrap()} and \texttt{facet\_grid()}) are adequate for
producing subplots of series and seasons. Additionally, the
\emph{sugrrants} package {[}49{]} provides a \texttt{facet\_calendar()}
function for creating calendar-style layouts of time series data. As
such, the {\emph{ggtime}} package does not currently implement any new
faceting functions, but a future direction for this work is the creation
of \texttt{facet\_cycle()}. As described in
Section~\ref{sec-future-grammar-components}, a time-aware facet function
would facilitate faceting by circular granularities without the data
pre-processing required by \texttt{facet\_wrap()} and
\texttt{facet\_grid()}.

\subsection{Limitations}\label{sec-grammar-limitations}

A primary design limitation of {\emph{ggtime}} is that while all
geometries in the ggplot2 ecosystem can use
\texttt{position\_time\_civil()}, only \texttt{geom\_time\_line()} is
able to fully leverage the offset aesthetics (\texttt{xtimeoffset} and
\texttt{ytimeoffset}) to correctly order and disambiguate repeated civil
times. We considered two approaches for addressing this limitation: (1)
extending additional geometries to be time-aware, and (2) implementing a
geometry-agnostic approach to handling offsets with a time coordinate
system. Neither approach adequately solved this limitation, since (1)
our design aims to leverage interoperability with the existing ecosystem
of {\emph{ggplot2}} extensions rather than replicating existing
functionality, and (2) visually appropriate decorations (e.g.~dashed
lines) to indicate jumps in time from changing offset vary by geometry.

\section{Visual design space}\label{sec-design-space}

As an extension of {\emph{ggplot2}}, there is a large design space of
temporal graphics that can be clearly expressed through combinations of
{\emph{ggtime}} grammar components, and an even larger space when
considering combined specifications with other compatible base and
extension {\emph{ggplot2}} objects. {\emph{ggtime}} not only improves
and shortens the specification of common static temporal graphics, but
the time-aware data and grammar components also increases friction for
creating semantically invalid temporal graphics.

\begin{figure}

\begin{minipage}{0.50\linewidth}

\centering{

\pandocbounded{\includegraphics[keepaspectratio]{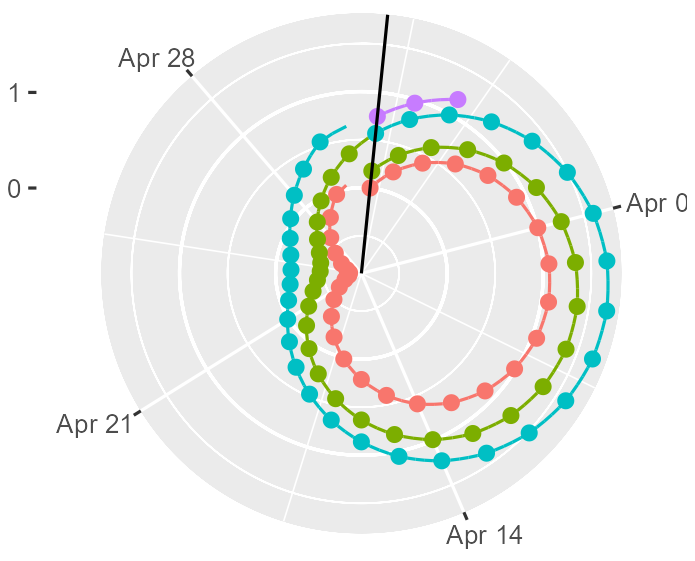}}

}

\subcaption{\label{fig-good-polar}Looped circular coordinates looping
over irregularly spaced time points produce ragged plots in polar
coordinate spaces.}

\end{minipage}%
\begin{minipage}{0.50\linewidth}

\centering{

\pandocbounded{\includegraphics[keepaspectratio]{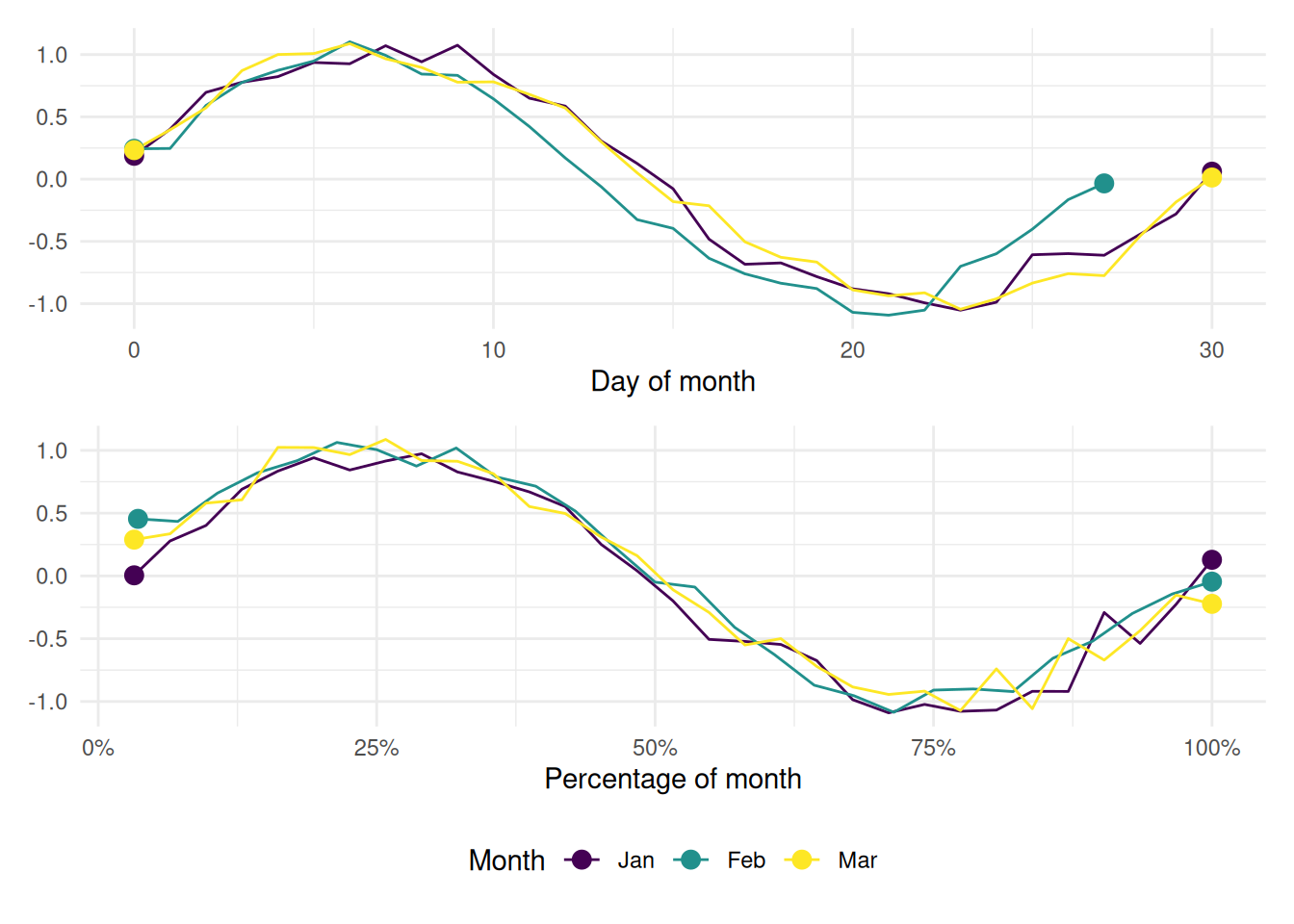}}

}

\subcaption{\label{fig-ragged-justified}Ragged time plots show data on a
time granule scale (e.g.~days), while justified plots show data on a
temporal progress scale (e.g.~\% of month).}

\end{minipage}%
\newline
\begin{minipage}{0.50\linewidth}

\centering{

\pandocbounded{\includegraphics[keepaspectratio]{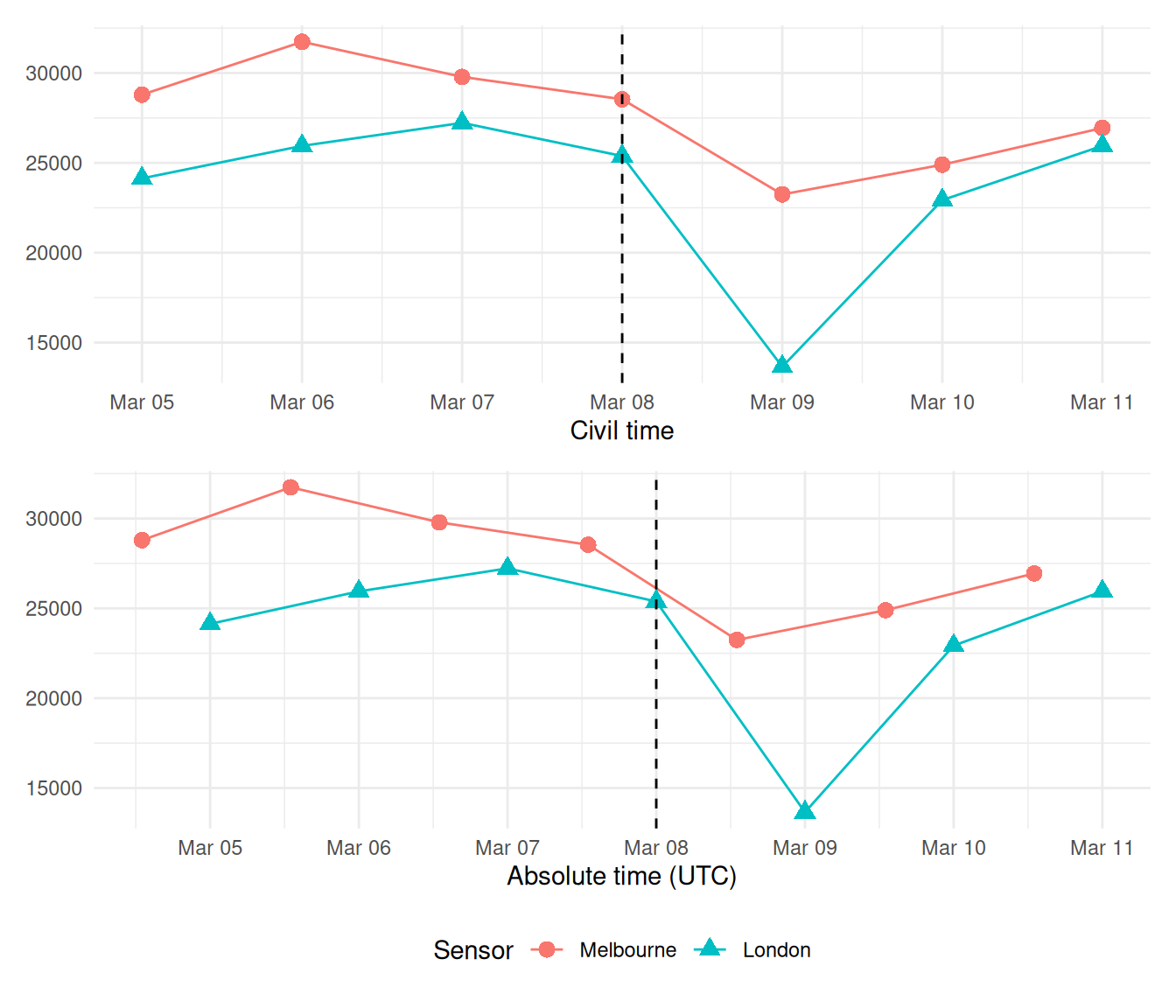}}

}

\subcaption{\label{fig-alignment}Alignment allows for observations from
different time zones to be shown according to civil time or absolute
time.}

\end{minipage}%

\caption{\label{fig-design-dense-support}Illustrative examples of how
different combinations of {\emph{ggtime}} grammar components can align
patterns over time.}

\end{figure}%

\subsection{Ragged and Justified Plots}\label{sec-design-space-ragged}

Different combinations of these grammatical elements produce different
styles of time plots that support the visualization of cyclical patterns
with irregular periodicity: {\emph{ragged}} and {\emph{justified}}
plots. The terms are analogous to text alignment styles, and describe
how time series data that are looped over circular coordinate systems
can exhibit different alignments based on how differences in the length
of a looping period are handled (i.e., monthly periodicity in the
Gregorian calendar is irregular).

In \emph{ragged} plots, loops over variable lengths of time without any
changes to the timescale are shown on a circular coordinate system. This
plot style is \emph{ragged} because the lengths of sub-intervals reflect
the length of absolute time of each cycle, allowing shorter cycles to
end earlier than longer cycles. Since each cycle shares the same
timescale, the overall length of the time axis reflects the length of
the longest cycle. For example, looping a time series by month produces
a timescale of 1--31 days, with months that have fewer than 31 days
ending early. If normalization is not applied to each loop, this results
in the visual effect of \emph{ragged} ends as shown in the top panel of
Figure~\ref{fig-ragged-justified}. Similarly, looping by day produces a
timescale of 1--25 hours, with days that have 23 or 25 hours (e.g.~due
to daylight saving transitions) causing subsequent days to be offset by
one hour. Ragged plots are useful when differences in the length of
cycles are meaningful, such as when tracking menstrual health.

This style of plot can be produced with {\emph{ggtime}} when using
\texttt{coord\_loop()} or \texttt{coord\_calendar()} with the
\texttt{loops} argument to specify a known time point in each cycle, or
with \texttt{time\_loops} to specify irregular calendar granularities
(e.g.~months).

The \emph{justified} plot style arises when circular coordinate systems
show normalized cycle lengths. Normalization refers to stretching or
compressing intervals of variable amounts of physical time (e.g.~number
of days in a month) to share the same effective length (e.g.~percentage
through month). The time-axis scale now represents fractions of the
normalized looping granule. For example, in a justified layout, if each
row in the layout is a month, then 0.1 on the time-axis represents 10\%
through the month, which is 3 days into a 30 day month but 2.8 days into
a non-leap year February. Justified layouts are helpful for identifying
shared or similar cyclical patterns or structures across cycles with
variable lengths, such as with sunspots and solar cycles.

In {\emph{ggtime}}, the timescale can be warped to normalize the length
of cycles using
\texttt{scale\_x\_mixtime(warps\ =\ \textless{}time\textgreater{})},
where \texttt{warps} specifies a known starting point for each cycle
(e.g.~the cycle's trough). Alternatively, the \texttt{time\_warps}
argument can be used to specify irregular calendar granularities
(e.g.~months). Combining this warped timescale with matching
\texttt{loops} or \texttt{time\_loops} for \texttt{coord\_loop()} or
\texttt{coord\_calendar()} produces justified plots.

\subsection{Absolute and Civil Time
Plots}\label{absolute-and-civil-time-plots}

Absolute and civil time plots are produced in {\emph{ggtime}} by using
\texttt{position\_time\_civil()} to plot time series observations by
their position in a shared `absolute' reference time (e.g.~UTC time, as
in the bottom panel of Figure~\ref{fig-alignment}), or relative to a
shared `civil' timescale (e.g.~the 24 hour clock, as shown in the top
panel of Figure~\ref{fig-alignment}). In contrast with
\texttt{ggplot::geom\_line()}, {\emph{ggtime}} defaults to civil time
positioning for \texttt{geom\_time\_line()}, producing civil time plots
rather than absolute time plots. The positioning of time series
observations according to civil time is particularly useful for
visualizing patterns in human activity, such as commuting patterns and
other daily routines.

\subsection{Other Parts of the Temporal Graphics
Landscape}\label{other-parts-of-the-temporal-graphics-landscape}

\begin{figure}

\centering{

\pandocbounded{\includegraphics[keepaspectratio]{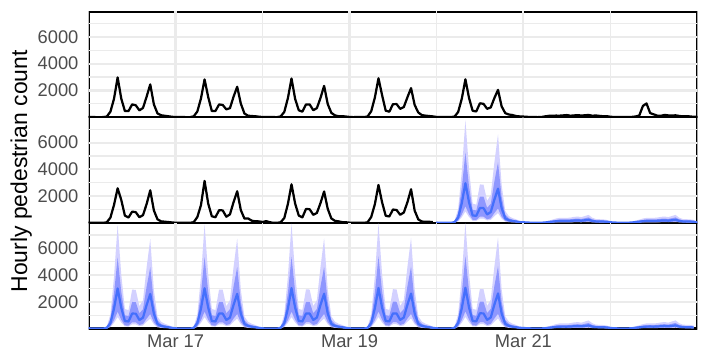}}

}

\caption{\label{fig-good-forecast}A calendar arrangement of forecasts
created by combining calendar coordinates from {\emph{ggtime}} with line
ribbon interval geometries from \emph{ggdist}.}

\end{figure}

When analyzing and visualizing datasets that incorporate multiple types
of semantics (e.g.~forecasts combine time and uncertainty), it may
initially seem necessary to define specific rules for combining these
domains in a single plot (e.g.~a forecast plot). However, the Grammar of
Graphics, and by extension, the layered grammar of graphics and
{\emph{ggplot2}}, are designed to support the integration of multiple
types of data semantics into a single statistical graphic.
Interoperability in \emph{Grammar of Graphics} systems simply requires
that there are valid mathematical interactions between each domain in
the combination, and that appropriately defined grammar components for
each domain implemented in a common system (e.g.~{\emph{ggplot2}}) will
integrate seamlessly. Forecasting, as the application of statistical
methods to time series data for the purposes of prediction, clearly
satisfies the first condition. The second condition is met in
{\emph{ggplot2}} by the \emph{ggdist} package {[}19{]}. As illustrated
in Figure~\ref{fig-good-forecast}, the combination of
\texttt{ggtime::coord\_calendar()} and
\texttt{ggdist::geom\_line\_ribbon()} produces a calendrical variation
on the standard linear forecast plot.

\section{Usage scenarios}\label{sec-usage-scenarios}

To demonstrate the utility of {\emph{ggtime}}, we present selected
visualization issues arising from incorrect or incomplete handling of
temporal semantics and show how {\emph{ggtime}} facilitates more nuanced
handling of these semantics.

\subsection{Rates of change in civil time}\label{sec-usage-slopes}

Often when answering questions about patterns in human behavior, the
model of time used in analysis will need to reflect the location, the
local calendar, and the local time zone of the data. For example, when
considering adjustments for daylight saving shifts, patterns of human
activity generally shift sharply to adhere to civil time
(e.g.~pedestrian foot traffic around a train station peaking at `8am'
before and after a daylight saving changeover). Similarly, sales data by
hour-of-the-day for stores across a large country, such as the United
States, will show different patterns depending on the time zone of the
store. By contrast, animal activity will not shift during DST
changeovers (since absolute time is unaffected), and electricity usage
may depend on the timing of sunset, which varies by location even within
the same time zone.

\begin{figure}

\centering{

\pandocbounded{\includegraphics[keepaspectratio]{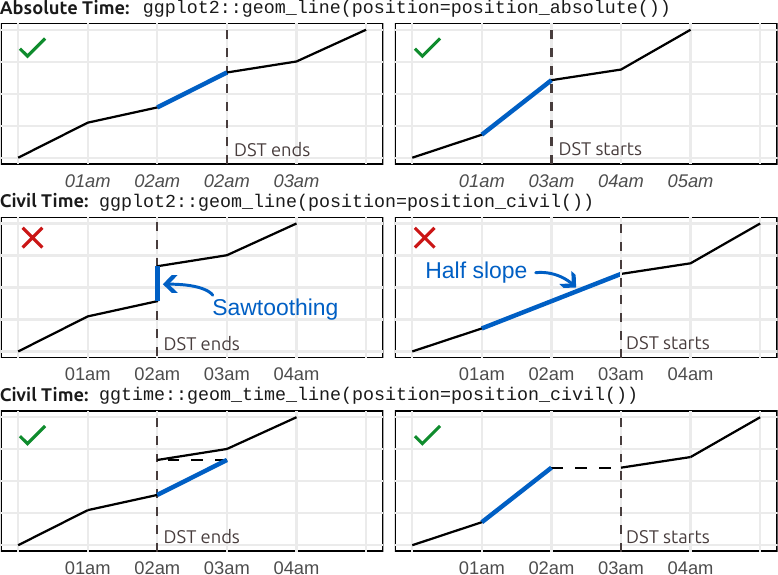}}

}

\caption{\label{fig-dst-slopes}A comparison of line geometries
positioned in absolute and civil time during a daylight saving time
transition. The common use of ggplot2's \texttt{geom\_line()} results in
inaccurate line slopes for data positioned in civil time.}

\end{figure}

Many locations observe daylight saving time (DST) changes twice per
year, shifting local civil time forward one hour in spring and backward
one hour in autumn. This results in discontinuities in civil time, which
can lead to misleading visualizations if not handled correctly. For
example, lines are commonly used in visualizing time-oriented data, and
their slopes indicate the rate of change in measurements over time.
Persistently increasing or decreasing lines indicate trends, while sharp
changes in the slope are associated with special events or data
anomalies. When visualizing data that span DST transitions, traditional
plotting libraries may produce incorrect slopes during these periods
because they do not account for the time shift.
Figure~\ref{fig-dst-slopes} compares the results obtained from ggplot2
and ggtime when plotting data that span a DST transition.

Converting from absolute time to civil time resolves the misalignment
resulting from time zone transitions, but this conversion is inherently
lossy, as multiple absolute time points map to the same civil time point
when clocks are set backward. Figure~\ref{fig-dst-slopes} shows how the
loss of temporal information produces inaccurate slopes, since the
length of time displayed on the x-axis no longer reflects the actual
amount of time that has passed. This can result in saw-toothing when
civil time repeats, and a flatter slope when civil time skips forward.

The {\emph{ggtime}} positional scales \texttt{position\_time\_civil()}
and \texttt{position\_time\_absolute()} remove the need for data
pre-processing and provide additional context via the
\texttt{{[}x/y{]}timeoffset} aesthetics for \texttt{geom\_time\_line()}.
This additional information disambiguates the ordering of time, allowing
the geometry to accurately represent and account for time zone
transitions using dashed lines parallel to the time axis. These time
offset aesthetics can also be used to visualize other sudden changes in
time, such as travel across time zones and machine clock adjustments.

\subsection{Common timescales for mixed
granularities}\label{sec-usage-nyt}

Data from multiple sources are often available at different
granularities, complicating visual comparisons. In such cases, it is
common to represent coarser granularities (e.g.~months) as the first
moment in a finer granularity (e.g.~the first day of the month).
However, this left alignment can be misleading and lead to incorrect
interpretations of the timing of events relative to coarser
granularities.

\begin{figure}

\centering{

\pandocbounded{\includegraphics[keepaspectratio]{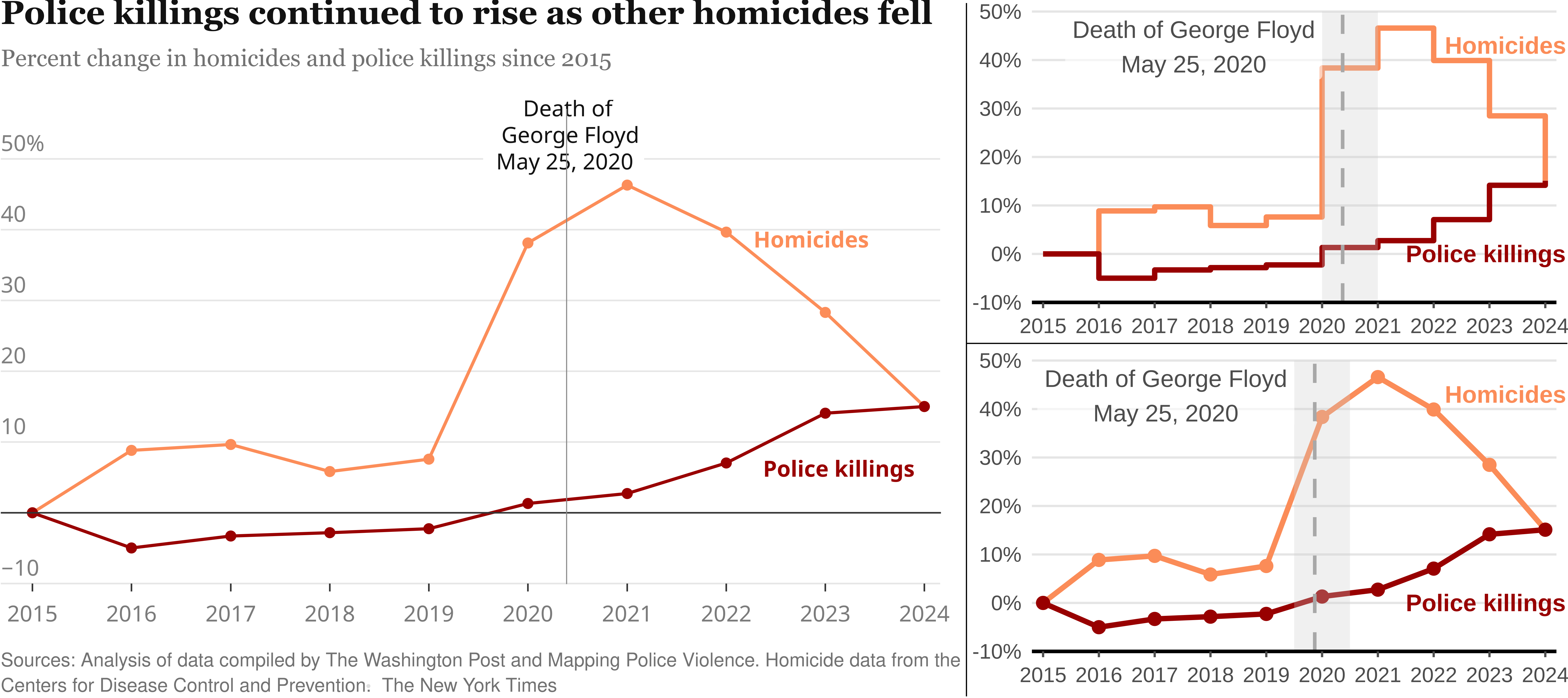}}

}

\caption{\label{fig-nyt-floyd}The left plot replicates a published
\emph{New York Times} figure which positions annual data at the first
day of each year, causing a potentially misleading interpretation of
changes around the time of the annotated date. The right two plots show
alternative visualizations that are less likely to mislead.}

\end{figure}

For example, consider Figure~\ref{fig-nyt-floyd}. The left plot
replicates a figure published by the \emph{New York Times} showing
annual homicides and police killings, with the date of George Floyd's
death. This figure uses left alignment of the coarser annual
granularities, which causes the alignment of the daily event to be
potentially misleading. Did homicides and police killings really
continue to rise unabated after the death of George Floyd?

Two alternative visualizations are presented on the right side of
Figure~\ref{fig-nyt-floyd}: the top right uses step geometries to
represent the available annual information as constant over the year,
while the bottom right centers each year's value at the middle day of
the year. Both approaches use more accurate handling of temporal
indeterminacy around the annotation, and neither suggests that the rate
of increase in homicides and police killings was unchanged after George
Floyd's death.

ggtime's \texttt{scale\_*\_mixtime()} defaults to center alignment of
coarser granularities when plotting mixed granularities. This alignment
can be changed to be any fractional part of granularity (0--1). This
process is equivalent to converting a discrete time model to a
continuous time model, whereby adding a fractional part to the coarser
granularity effectively identifies an exact moment in continuous time.

\subsection{Navigating linear and cyclical arrangements of
time}\label{sec-usage-circular}

Different arrangements of time reveal different patterns in
time-oriented data. Linear arrangements of time are useful for showing
changes over time (e.g.~trends), while cyclical arrangements of time are
useful for showing repeating seasonal patterns within cyclic
granularities. The latter use circular or looped layouts to facilitate
comparisons across granularities, such as months, weeks or days. It is
common in an analysis to switch between linear and multiple cyclical
arrangements of time to explore different aspects of the data.

For example, {[}23{]} shows painting sales data by day of the week, and
faceted by month of the year, to reveal weekly and seasonal patterns in
sales. Another example is the visualization of pedestrian counts by time
of day and day of the week in Figure~\ref{fig-good-calendar}. In each
case, substantial pre-processing of the data is required to convert from
the original linear arrangement of time to the desired cyclical
arrangement. This pre-processing is often cumbersome and error-prone,
and can lead to a loss of temporal semantics (e.g.~ordering of days of
the week, ordering of months, length of months, the periodic nature of
circular granularities, etc.). Even if the temporal semantics are
preserved, the code is often unnaturally complex and difficult to read.

\begin{figure}

\centering{

\pandocbounded{\includegraphics[keepaspectratio]{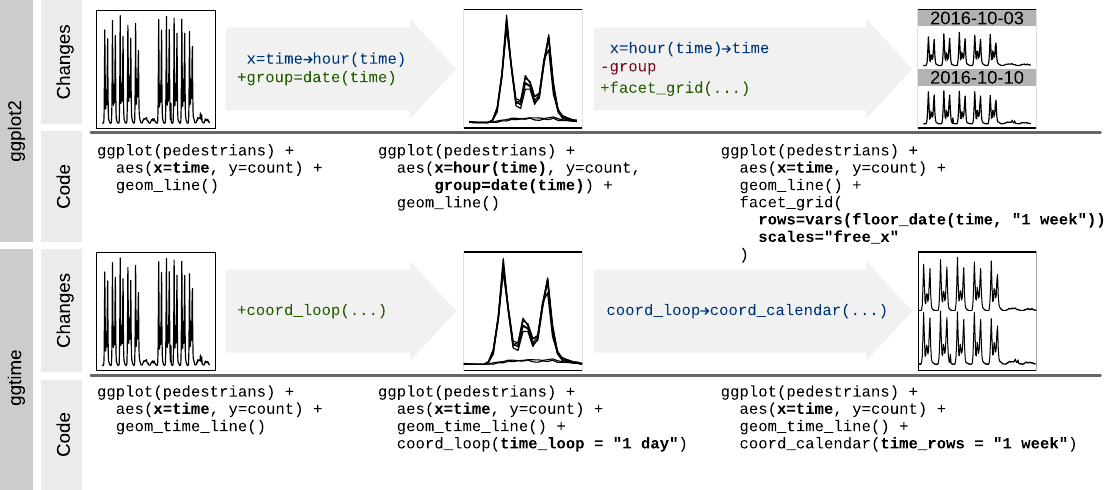}}

}

\caption{\label{fig-code-coordinates}A comparison of code used in
navigating from linear to cyclical {\emph{arrangements}} of temporal
visualization. This is commonly achieved in {\emph{ggplot2}} with data
pre-processing in aesthetics or with facets and discretization of time;
both approaches are cumbersome and lossy. The equivalent code in
{\emph{ggtime}} is shorter and more consistent, where linear data
{\emph{arrangements}} are combined with either linear or cyclical
coordinate spaces.}

\end{figure}

Figure~\ref{fig-code-coordinates} shows code for the visual comparison
of hourly pedestrian counts across 14 days, using {\emph{ggplot2}} and
{\emph{ggtime}}. The data are shown plotted against time in the left
panel, against time of day in the middle panel, and against time of week
with faceted weeks in the right panel. Even with the helper functions
\texttt{hour()}, \texttt{date()} and \texttt{floor\_date()}, the
{\emph{ggplot2}} code is more opaque than the {\emph{ggtime}}
equivalent, which uses \texttt{geom\_time\_line()}, and the coordinate
systems \texttt{coord\_loop()} and \texttt{coord\_calendar()}.

This figure illustrates how {\emph{ggtime}} facilitates fluid navigation
between linear and cyclical arrangements of time by switching coordinate
systems, while maintaining temporal semantics through the use of
cyclical arrangements of linear granularities. The ability to quickly
switch between multiple cyclical arrangements of time reduces friction
when visualizing repeating patterns across multiple cyclical
granularities. The above example shows how to switch between daily and
weekly arrangements of time, which is also commonly needed for annually
repeating patterns.

In addition to the simpler interface, the preservation of complete
temporal semantics using linear time offers several benefits. While not
shown in Figure~\ref{fig-code-coordinates}, the labeling of axes and
legends is also simplified in {\emph{ggtime}} by using scales that are
aware of the temporal semantics of the data, rather than relying on
manual specification of breaks and labels for each arrangement of time.
The coordinate systems are also more accurate, allowing the positioning
of observations to reflect the actual length of time intervals
(e.g.~months with 28, 30 and 31 days). Since time is treated linearly,
the end of one line is directly connected to the start of the next line,
and geometries and statistics that span across cycles (e.g.~smoothing)
are correctly calculated.

\section{Discussion}\label{sec-discussion}

\subsection{Future directions}\label{sec-future-grammar-components}

There are several additional temporal graphics that are commonly used in
practice but are not yet directly implemented in {\emph{ggtime}}. Many
of these graphics can be constructed from combinations of existing
{\emph{ggtime}} and {\emph{ggplot2}} components, but creating time-aware
variants of these functions would simplify their use and ensure they
respect temporal semantics.

As described in Section~\ref{sec-grammar-geometries}, {\emph{ggtime}}
currently provides a single time-aware line geometry
(\texttt{geom\_time\_line()}) which better preserves temporal semantics
to maintain accurate slopes (Section~\ref{sec-usage-slopes}). A select
few additional geometries commonly used in time series visualization
would be useful additions to {\emph{ggtime}}, including area geometries
for stacked time series plots and labels for visualizing time intervals.

\begin{figure}

\begin{minipage}{0.50\linewidth}

\centering{

\includegraphics[width=0.9\linewidth,height=\textheight,keepaspectratio]{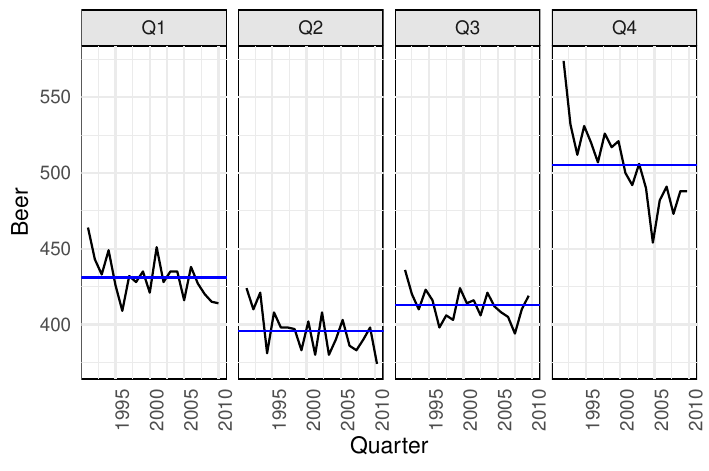}

}

\subcaption{\label{fig-good-subseries}A linear time plot faceted by
cyclical granularities provides an alternative approach to visualizing
repeating patterns. Different line slopes indicate that the pattern is
changing over time.}

\end{minipage}%
\begin{minipage}{0.50\linewidth}

\centering{

\includegraphics[width=0.9\linewidth,height=\textheight,keepaspectratio]{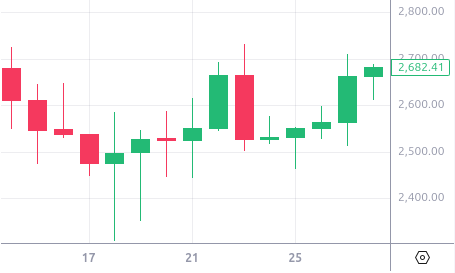}

}

\subcaption{\label{fig-good-candlestick}A candlestick chart summarizing
intraday variability of a stock. The lines indicate the highest and
lowest prices, while the box shows the opening and closing price.}

\end{minipage}%

\caption{\label{fig-future-work-temp}Two charts that could be
implemented more easily in future versions of {\emph{ggtime}}.}

\end{figure}

The candlestick chart shown in Figure~\ref{fig-good-candlestick} is
another worthwhile future direction. Candlestick charts are commonly
used in time series analysis for financial analysis to show summaries of
price movements over temporal aggregations. This geometry would require
both a statistical transformation (e.g.~\texttt{stat\_ohlc()}) and a
geometry (e.g.~\texttt{geom\_time\_candlestick()}) to implement. More
general statistics over time would also be useful, such as sliding
window statistics (e.g.~rolling averages) and time-based aggregations
(e.g.~daily averages).

The faceting capabilities of ggplot2 are adequate for separately
visualizing categories of time (e.g.~weekdays and weekends), however,
they require data pre-processing that is prone to errors since it
requires discretizing time. A time-aware variant for faceting would
simplify this process and can be helpful in arranging time to more
clearly show patterns both linearly and cyclically. Cyclical
arrangements of facets reveal changes in seasonal patterns over time, as
is shown in Figure~\ref{fig-good-subseries}. To better facilitate this
type of visualization, a \texttt{facet\_cycle()} function could be
implemented to facet by cyclical granularities (e.g.~day of week, month
of year) while preserving the correct ordering of the facets.

The geoms provided by {\emph{ggtime}} cover many common temporal
visualization needs, but there are some additional extensions and
enhancements that could be explored in future work. For example,
financial time series visualizations often include specialized summary
geometries such as candlestick charts, depicting open/high/low/close
prices over a time interval, which could be implemented as additional
geoms or statistical transformations.

We plan to support durations and intervals more fully in future work,
potentially through a \texttt{geom\_span()} that can visually encode
time spans as horizontal bars, using \texttt{xmin} and \texttt{xmax}
aesthetics to represent the start and end of the interval. This would
allow for effective visualization of events or periods defined by
durations or intervals, complementing the existing point-based temporal
geoms.

The principles and design patterns established in {\emph{ggtime}} can
serve as a foundation for future work in visualization systems for other
domains, where semantic integrity can be lost using general-purpose
visualization tools. By formalizing the representation and encoding of
temporal data, {\emph{ggtime}} provides a model for how other semantic
domains can be similarly supported through grammar-based visualization
systems.

Future work on semantic visualization systems, including
semantic-preserving extensions to graphical grammars, will likely also
result in new research and evaluation methods. For instance, although
corpora of visualization datasets and/or graphics collected from current
and archival sources (e.g.~VizNet {[}14{]}; ZuantuSet {[}27{]}) have
many useful research applications, as samples they are fundamentally
unable to fully characterize semantically valid visualization design
spaces. Nevertheless, attempts to characterize semantic domains and
validity could be informed by task-focused abstractions in existing
visualization design studies {[}29,42{]}.

\subsection{Dialogues between statistical graphics and
InfoVis}\label{dialogues-between-statistical-graphics-and-infovis}

In earlier drafts of this paper, there was some back-and-forth between
the authors (whose backgrounds vary from statistics to information
visualization and human-computer interaction) regarding whether the
appropriate scope of our work is restricted to exploratory statistical
graphics or if it can be extended to communicative visualizations as
well. Through a number of somewhat circuitous discussions, it was
discovered that the authors with a statistical background somewhat
conflated `communication design' with `visual design', and consequently
were hesitant to claim contributions towards communicative
visualizations. We were eventually able to articulate and confirm a
mutually shared view that the temporal graphics generated in an EDA
workflow to \emph{search for statistical insights}, and graphics created
to accurately encode and \emph{communicate statistical insights},
usually need to satisfy the same set of semantic-preserving conditions.
Furthermore, the design requirements and evaluation criteria for a tool
which allows one to navigate the space of semantic-preserving temporal
encodings are essentially identical---while audiences differ, and
communicative visualization design typically involves other
considerations (e.g.~how a visualization uses text or visual hierarchy
to guide reading, how it integrates into a surrounding communication
context, etc.), in both cases one would want to maintain semantic
validity while iterating through encoding designs. Indeed, visualization
grammars such as ggplot2 are often used by journalists when constructing
visualizations, a testament to the value of these tools in communicative
practice.

This misunderstanding might also offer some insight into the varied
usage of the term {\emph{`statistical graphics'}}. This term has been
used, often by statisticians {[}6,10,52{]}, to gesture towards some
seemingly definite, yet still elusive, distinctions between
visualizations produced by statisticians relative to other fields such
as information visualization. These gestures are often accompanied by
disclaimers about how statistical graphics are not meant to be `pretty'
or `polished' {[}45,56{]}, which likely reflect similar
misunderstandings about the role of `visual design' held by authors
prior to this collaboration. On the other hand, the sense in which
`statistical graphics' is used in Information Visualization and related
fields can be somewhat narrowly defined via example rather than from
first principles. For example, in {[}22{]}, the authors conduct a case
study on their metric-based evaluation method for visualization
notations in the context of `conventional {[}static{]} statistical
graphics'. As explained by the authors, their gallery of examples was
designed to cover a wide variety of commonly used chart forms reflective
of the domain of interest\footnote{``for instance, in statistical
  graphics, how heat maps or histograms are produced'' {[}22{]}}, with
input from their expert interviewees. This approach appears grounded in
a desire for ecological validity, which is in many contexts a useful and
appropriate validation approach. However, this somewhat more narrow
notion of `statistical graphics' could also contribute to the frictions
in dialogue between statisticians, especially those researching new
statistical methods and accompanying novel statistical graphics, and the
InfoVis community.

Resolving these mutual misunderstandings---seeing past narrow,
surface-level views of either field---we discovered a shared goal in
building ways to effectively navigate spaces of semantically valid
visualization encodings. This goal has wide interest in both the
statistical graphics and information visualization communities. We hope
to continue building bridges across our communities to work towards this
larger vision.

\section{Conclusion}\label{sec-conclusion}

This work illustrates that once semantic-validation conditions have been
defined, a graphical grammar can be designed to facilitate fluid
iteration over visualizations that satisfy these formal principles and
conditions, whilst avoiding incorrect or misleading visualization
designs.

We have also illustrated the utility of graphical grammars that maintain
domain specific notions of validity, and have shown that they help
reduce visualization mistakes, and make iterating over the space of
semantically-valid graphics more fluid. We have presented formal
semantic validation conditions for temporal graphics based on temporal
semantics relevant to time series analysis and visualization.

These temporal semantics and accompanying conditions have been
translated into {\emph{ggtime}}, a grammar-based system that builds on
the popular {\emph{ggplot2}} visualization package in R. We have
designed time-aware data structures to represent temporal data, created
a layered set of time-aware geoms, and provided temporal coordinate
systems, enabling users to easily compose semantically-valid temporal
graphics, and smoothly navigate the design space of existing and novel
temporal graphics. By enforcing semantic validity, {\emph{ggtime}} helps
prevent common pitfalls in temporal visualization, and facilitates more
effective and fluid exploratory data analysis.

The expressive grammar of {\emph{ggtime}} allows for a wide range of
temporal visualizations, from simple time series plots to more complex
cyclical and seasonal representations. The grammar-based approach of
{\emph{ggtime}} allows for a flexible and extensible framework that can
be used to develop new temporal analysis tasks, and is interoperable
with other semantic variables. The integration with the {\emph{ggplot2}}
ecosystem ensures that users can leverage existing tools and workflows
while benefiting from the enhanced temporal capabilities provided by
{\emph{ggtime}}.

While there have been previous efforts to incorporate temporal semantics
into visualization systems, {\emph{ggtime}} is distinguished as a
grammar-based approach that ensures semantic validity throughout the
visualization process, addressing both data representation and visual
encoding. This is in contrast to other systems that may rely on data
preprocessing or specialized visual encodings without a formal grammar
underpinning, or which are closely tied to specific application domains
or languages.

\section*{References}\label{references}
\addcontentsline{toc}{section}{References}

\phantomsection\label{refs}

\end{document}